\documentclass[prb,twocolumn,showpacs,amsmath,amssymb,superscriptaddress,floatfix]{revtex4}
\pdfoutput=1

\usepackage{ graphicx } 
\usepackage{ bm } 
\usepackage{ color }

\newcommand{\be}{\begin{equation}}
\newcommand{\ee}{\end{equation}}
\newcommand{\bea}{\begin{eqnarray}}
\newcommand{\eea}{\end{eqnarray}}

\begin{document}
\title{Edge mode dynamics of quenched topological wires}
\author{ P. D. Sacramento }
\affiliation{ \textit CeFEMA,
Instituto Superior T\'ecnico, Universidade de Lisboa, Av. Rovisco Pais, 1049-001 Lisboa, Portugal }

\date{ \today }


\begin{abstract}
The fermionic and Majorana edge mode dynamics of various topological systems is compared, 
after a sudden global quench of the Hamiltonian
parameters takes place. Attention is focused on the regimes where the survival probability
of an edge state
has oscillations either due to critical or off-critical quenches.
The nature of the wave functions and the overlaps between the eigenstates of different
points in parameter space determine the various types of behaviors, and the distinction
due to the Majorana nature of the excitations plays a lesser role. Performing a sequence
of quenches it is shown that the edge states, including Majorana modes, may be switched
off and on. Also, the generation of Majoranas due to quenching from a trivial phase
is discussed.
\end{abstract}


\maketitle

\section{Introduction}

Quenching a quantum system raises interesting questions \cite{eisert},
particularly when the
evolution is unitary. When a sudden quench takes place the
evolution is determined by the overlaps between the eigenstates
of the instantaneous Hamiltonians, prior and after the quench,
expressed by a given change of a set of parameters on which the
Hamiltonian depends.

An abrupt change of the state of an isolated quantum system leads to a unitary
time evolution and, therefore, the issue of thermalization has been addressed
\cite{polkovnikov1,exp}. 
In general, it is expected 
that correlation functions stabilize \cite{thermal1,thermal2,thermal3,thermal4,thermal5,thermal6,thermal7,thermal8,thermal9,thermal10,thermal11,thermal12,thermal13,thermal14,thermal15,thermal16,thermal17,thermal18,deutsch,rigol1}.
In the cases of soluble and integrable systems thermalization breaks down as one
approaches an integrable point. However, some sort of 
thermalization is predicted
for which an equilibrium like distribution is expected in terms of
a generalized Gibbs ensemble, of the (infinitely) many conserved
quantities \cite{rigol2,cazalilla1,cazalilla2,cazalilla3,cazalilla4,cazalilla5,cazalilla6,cazalilla7,cazalilla8,cazalilla9,cazalilla10,cazalilla11,cazalilla12,cazalilla13,cazalilla14,cazalilla15,cazalilla16,cazalilla17,rigol3,luttinger}.

An interesting case is the effect of a sudden quench of the
parameters of an Hamiltonian with topological properties
and, particularly, a change of parameters that leads to a change
of topological properties. Specificaly, how the 
topological properties and
topological edge
states respond to such quenches and how robust they are.
Topological systems have attracted interest \cite{kane,zhang} and, specifically,
topological superconductors \cite{alicea} due to the prediction of Majorana fermions
\cite{mourik,bernevig1,science,glazman,dassarma}.
It has been shown before that topological systems are quite robust to a 
quantum quench, as exemplified by the toric code model \cite{tsomokos,hamma}. 

It has also been shown recently \cite{pre} that,
in the case of quenches in infinite size topological superconductors,
the Chern number can not be changed by a unitary
evolution. The same result was shown more generally for any topological system, 
even though it is possible to change the Bott index and topology 
if the system has a finite size \cite{alessio}.
Therefore, in a finite system the Chern number may change \cite{pre} and
the response of the edge states to a time dependent perturbation in finite systems 
may not be protected by topology.
Furthermore, quenches in superconducting systems with topological properties, 
performed self-consistently \cite{dzero3}, showed the importance of the topological
properties in the evolution of the system \cite{sen1} and raised questions regarding the
survival of the topological order to the quench \cite{tsomokos,scheurer,dzero4,dzero5}. 
So the issue is not resolved and is attracting considerable attention.

In general, quenches that lead the system from a topological
phase to a trivial phase imply a decay of the gapless edge
states and, in the reverse quenching, the topological states
are not generated. However, since the systems have finite extent,
a revival of the original states is observed with a revival time
that scales with the system size. Also, as will be shown here, Majorana
edge states may be generated from a trivial phase under appropriate conditions.

The behavior of edge states under an abrupt quantum quench has been considered
very recently in the context of a two-dimensional topological insulator \cite{bhz},
where it was found that, in the sudden transition from the topological insulator to
the trivial insulator phase, there is a collapse and revival of the edge states \cite{patel}.
Similar results were obtained for the one-dimensional Kitaev model \cite{rajak}, also
studying the signature of the Majoranas in the entanglement spectrum \cite{chung}.
Their dynamical formation and manipulation has been considered in \cite{perfetto} and
\cite{scheurer}. The robustness of edge states may also be studied in the context of
slow quenches from a topological phase to a trivial phase. 
\cite{bermudez1,bermudez2,pre}. 

The effect of parity blocking on the dynamics of the edge modes has been considered recently
in which case the dynamics is restricted if there is a change in fermion parity accross
the quench \cite{blocking}.
On the other hand, the Majorana zero modes lead to some universal non-equilibrium signature
in the Loschmidt echo with an universal exponent associated with the algebraic decay
\cite{moore1}. Also, the dynamics of the tunneling into non-equilibrium edge states has been
proposed as a possible signature of the existence of these states \cite{moore2}.
Non-equilibrium situations also may allow the transport of Majorana edges states using extended
gapless regions with a small but finite overlap with the Majoranas \cite{dutta1}. Their effect
has also been considered in \cite{sen2} and in \cite{zvyagin}.

While quenches from a topological to a trivial phase lead to decay of the edge states,
quenches within the same topological phase typically lead to a decrease of
the revival probability, but to a finite value. 
An intermediate case of a quench
between two topologically different phases and within the same phase is
a quench to a critical point or critical line, separating two different topological phases.
At this point the spectrum becomes gapless. In some appropriate conditions 
the survival probability shows oscillations \cite{rajak}. 
In this work particular attention
will be paid to these oscillations in various topological systems.

We will consider sudden quantum quench in one-dimensional systems and focus on 
the oscillations at critical quenches and non-critical quenches.
The survival probability will be studied taking into account finite size effects.
A period doubling is observed as the initial state of the quench is moved sufficiently
far from the critical region.
The dynamics of Majorana zero energy states is compared with that of finite-energy excited states and
the dynamics of Majoranas is also compared with that of fermionic zero energy modes. 
The importance of overlaps and spectrum of the final eigenstates is shown.
It is shown that an appropriate sequence of quenches may lead to a on/off process
of the existence of Majorana states, with potential application in their manipulation
and storing of information.
Also, it is shown that states of a trivial phase may lead after a quench to a
topological phase to Majorana states.

Universal single-frequency oscillations in the entanglement spectrum of two Kondo
impurities coupled to leads have been discovered recently \cite{henrik}.
These are the result of a off-critical quench accross the phase transition to a Kondo
screened regime where all the spins are coupled to the impurity spins.
The frequency scales with the inverse of the system size.
In the systems considered here the oscillations in general are a 
superposition of different frequencies.

The paper is organized as follows. In section II a discussion of the quantities
calculated after a sudden quench of the parameters of the Hamiltonian is presented.
In section III the topological models considered are briefly discussed.
In section IV the dynamics of the edge modes of Kitaev's model is studied
for the case of a single quench and the generation of Majorana modes
due to a quench, or sequence of quenches, is discussed. In section V the dynamics
of the edge modes of some multiband systems is discussed and the dynamics
of Majorana modes and fermionic zero energy modes are compared.
In section VI the conclusions are presented.

\section{Single-particle states and quantum quenches}

Let us consider an Hamiltonian defined by an initial set of
parameters $\xi_0$ for times $t<0$.
The single-particle eigenstates of the Hamiltonian are given by
\be
H(\xi_0) |\psi_{m_0}(\xi_0)\rangle = E_{m_0}(\xi_0)|\psi_{m_0}(\xi_0)\rangle .
\ee
At time $t=0$ a sudden transformation of the parameters is performed,
$\xi_0 \rightarrow \xi_1$. The eigenstates of the new Hamiltonian are given by
\be
H(\xi_1) |\psi_{m_1}(\xi_1)\rangle = E_{m_1}(\xi_1)|\psi_{m_1}(\xi_1)\rangle .
\ee
The time evolution of a single-particle state, $m_0$, is given by
\be
|\psi_{m_0}^I(t) \rangle = \sum_{m_1} e^{-i E_{m_1}(\xi_1) t} |\psi_{m_1}(\xi_1) \rangle
\langle \psi_{m_1}(\xi_1)|\psi_{m_0}(\xi_0) \rangle
\ee
for times $t \geq t_0$ (where $t_0=0$).
The survival probability of the initial state $|\psi_{m_0}(\xi_0) \rangle$ is, as usual,
defined by
\be
P_{m_0}(t)=|\langle \psi_{m_0}(\xi_0) |\psi_{m_0}^I(t) \rangle |^2
\ee

We may as well consider further quenches defined in a sequence of times and sets
of parameters as $t_0<t_1<t_2<t_3<\cdots$ and $\xi_0, \xi_1,\xi_2,\xi_3,\cdots$,
respectively. These intervals define regions as
$I (t_0 \leq t <t_1), II (t_1 \leq t < t_2), III (t_2 \leq t <t_3), \cdots$.
The case of a single quench is clearly obtained taking $t_1 \rightarrow \infty$, and
so on for further quenches.

Consider now a case for which we have two quenches in succession. In this case we have that
\bea
|\psi_{m_0}^{II}(t) \rangle &=& e^{-i H(\xi_2)t} |\psi_{m_0}^I(t_1) \rangle \nonumber \\
&=& \sum_{m_2} e^{-i E_{m_2}(\xi_2)t} |\psi_{m_2}(\xi_2) \rangle \langle \psi_{m_2}(\xi_2)|
\psi_{m_0}^I(t_1) \rangle \nonumber \\
&=& \sum_{m_2} \sum_{m_1} e^{-i E_{m_2}(\xi_2)t} e^{-i E_{m_1}(\xi_1)t_1} |\psi_{m_2}(\xi_2) \rangle
\nonumber \\
& & \langle \psi_{m_2}(\xi_2)|\psi_{m_1}(\xi_1) \rangle \langle \psi_{m_1}(\xi_1)|\psi_{m_0}(\xi_0) \rangle
\eea

Choosing $\xi_2=\xi_0$ we get that for $t_1 \leq t<\infty$ ($t_2 \rightarrow \infty$)
the overlap with an initial state, $n_0$, is given by
\bea
\langle \psi_{n_0}(\xi_0)|\psi_{m_0}^{II}(t) \rangle &=& \sum_{m_1} e^{-i E_{n_0}(\xi_0)t}
e^{-i E_{m_1}(\xi_1) t_1} \nonumber \\ 
& & \langle \psi_{n_0}(\xi_0)|\psi_{m_1}(\xi_1) \rangle \langle \psi_{m_1}(\xi_1)|\psi_{m_0}(\xi_0) \rangle
\nonumber \\
& & 
\eea
Therefore, the probability to find a projection to an initial state, $n_0$, given that the
initial state is $m_0$ is given by
\bea
P_{n_0m_0}(t) &=& |\langle \psi_{n_0}(\xi_0)|\psi_{m_0}^{II}(t) \rangle |^2 \nonumber \\
&=& | \sum_{m_1} e^{-i E_{m_1}(\xi_1) t_1} \nonumber \\
& & \langle \psi_{n_0}(\xi_0)|\psi_{m_1}(\xi_1) \rangle \langle \psi_{m_1}(\xi_1)|\psi_{m_0}(\xi_0) \rangle |^2 ,
\nonumber \\
& &
\eea
which is independent of time.

We may now at some given finite time, $t_2$, change the parameters from $\xi_2 \rightarrow \xi_3$.
As before we may now find that for $t_2 \leq t<\infty$ the same probability is given by
\be
P_{n_0m_0}(t) = |\langle \psi_{n_0}(\xi_0)|\psi_{m_0}^{III}(t) \rangle |^2
\ee
where
\be
|\psi_{m_0}^{III}(t) \rangle = e^{-i H(\xi_3)t} |\psi_{m_0}^{II}(t_2) \rangle
\ee
The probability is now a function of time, $t$.

In this work only unitary evolution of single-particle states is considered
and effects of dissipation are neglected.

\section{Models}

Various models have topological properties and, due to the bulk-edge correspondance,
have gapless edge modes. Here we consider a few one-dimensional systems with non-trivial
topological properties.

\subsection{One-band superconductor: the Kitaev model}

\begin{figure}
\includegraphics[width=0.75\columnwidth]{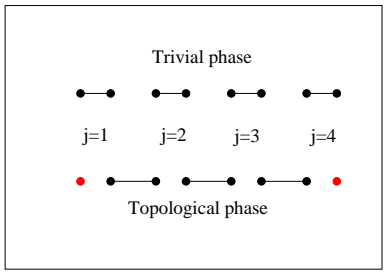}
\caption{\label{fig1}
(Color online) 
Phases of Kitaev model. 
At each lattice site, $j$, the two dots represent the two Majorana operators, $\gamma_{j,1}$ and
$\gamma_{j,2}$ (real and imaginary parts of $c_j$). The lines represent the links between
Majorana operators at given points in parameter space. 
Trivial phase with parameters: $\Delta=0, |\mu|>2t$ and topological
phase with $\mu=0, \Delta=t$.
In the trivial phase the two Majoranas at each site are linked in the Hamiltonian
and they constitute usual fermionic modes. In the topological phase the Majoranas are linked
at nearest-neighbor sites and the first and last Majorana operators are decoupled and therefore
have zero energy.
}
\end{figure}

The Kitaev one-dimensional superconductor with triplet p-wave pairing is described
by the Hamiltonian \cite{kitaev}
\bea
H &=& \sum_{j=1}^{\bar{N}} \left[ -t \left( c_j^{\dagger} c_{j+1} + c_{j+1}^{\dagger} c_j \right)
+ \Delta \left( c_j c_{j+1}+c_{j+1}^{\dagger} c_j^{\dagger} \right) \right] \nonumber \\
& - & \sum_{j=1}^N \mu \left( c_j^{\dagger} c_j -\frac{1}{2} \right)
\eea
where $\bar{N}=N$ if we use periodic boundary conditions (and $N+1=1$) or
$\bar{N}=N-1$ if we use open boundary conditions. $t$ is the hopping amplitude
taken as the unit of energy.
Using a Jordan-Wigner transformation defined by 
\bea
c_j &=& e^{i \pi \sum_{l=1}^{j-1} S^+(l) S^-(l)} S^-(j) \nonumber \\ 
c_j^{\dagger} &=& S^+(j) e^{-i \pi \sum_{l=1}^{j-1} S^+(l) S^-(l)} 
\eea
this model is equivalent to a spin-$1/2$ model
\bea
H &=& -\frac{1}{4} \sum_{j=1}^N \left[ \left(J_X-J_Y\right) \left( S^+_{j} S^+_{j+1}+
S^-_j S^-_{j+1} \right) \right. \nonumber \\
&+& \left. \left(J_X+J_Y \right) \left( S^+_{j} S^-_{j+1}+ S^-_j S^+_{j+1} \right) \right]
\nonumber \\
&-& \sum_{j=1}^N h_z S^z_j
\eea
The connection between the two models satisfies
\bea
t &=& \frac{1}{2} \left( J_X+J_Y \right) \nonumber \\
\Delta &=& \frac{1}{4} \left( J_X-J_Y \right) 
\eea
and the chemical potential is the magnetic field along $z$.
 
In momentum space the model is simply written as
\begin{eqnarray}
\hat H = \frac 1 2\sum_k  \left( c_k^\dagger ,c_{-k}   \right)
H_k
\left( \begin{array}{c}
c_{k} \\  c_{-k}^\dagger  \end{array}
\right)
\end{eqnarray}
where
\begin{eqnarray}
H_k = \left(\begin{array}{cc}
\epsilon_k -\mu & i \Delta \sin k \\
-i \Delta \sin k & -\epsilon_k +\mu  \end{array}\right)
\nonumber \\
\end{eqnarray}
with $\epsilon_k=-2t \cos k$.

A fermion operator may be writen in terms of two hermitian operators in the
following way
\bea
c_{j,\sigma} &=& \frac{1}{2} \left( \gamma_{j, \sigma, 1} + i \gamma_{j, \sigma, 2} \right) \nonumber \\
c_{j,\sigma}^{\dagger} &=& \frac{1}{2} \left( \gamma_{j, \sigma, 1} - 
i \gamma_{j, \sigma, 2} \right) 
\eea
The index $\sigma$ represents internal degrees of freedom of the fermionic operator, such as spin
and/or sublattice index, and the $\gamma$ operators are hermitian and satisfy the anti-commutation relations
\be
\{\gamma_m,\gamma_n \}=2 \delta_{nm}
\ee
In the case of the Kitaev model it is enough to consider $c_j=(\gamma_{j,1}+i \gamma_{j,2})/2$, since
the fermions are spinless.
In terms of these hermitian (Majorana) operators we may write that the Hamiltonian is given by,
using open boundary conditions,
\bea
H &=& \frac{i}{2} \sum_{j=1}^{N-1} \left[ 
(-t+\Delta) \gamma_{j,1} \gamma_{j+1,2}
+ (t+\Delta) \gamma_{j,2} \gamma_{j+1,1} \right] 
\nonumber \\
&-& \frac{i}{2} \sum_{j=1}^N \mu \gamma_{j,1}
\gamma_{j,2}
\eea

The chemical potential term involves all Majorana operators. Taking $\mu=0$ and selecting the
special point $t=\Delta$ the Hamiltonian reduces considerably to
\be
H(\mu=0, t=\Delta) = it \sum_{j=1}^{N-1} \gamma_{j,2} \gamma_{j+1,1}
= -it \sum_{j=1}^{N-1} \gamma_{j+1,1} \gamma_{j,2} 
\ee
It is easily seen that the operators $\gamma_{1,1}$ and $\gamma_{N,2}$ are missing
from the Hamiltonian. Therefore there are two zero energy modes. Defining from
these two Majorana fermions a single usual fermion operator (non-hermitian)
taking one of the Majorana operators as the real part and the other as the imaginary
part, its state may be either occupied or empty with no cost in energy.
Defining $d_j=1/2 \left( \gamma_{j,2}+i \gamma_{j+1,1} \right)$ and
$d_N=1/2 \left(\gamma_{N,2}+i \gamma_{1,1} \right)$ we can write the Hamiltonian as
\be
H=t \sum_{j=1}^{N-1} \left( 2 d_j^{\dagger} d_j -1 \right) + \epsilon_N \left(2 d_N^{\dagger} d_N
-1 \right)
\ee
with $\epsilon_N=0$. Therefore the fermionic mode $d_N$ does not appear in the Hamiltonian
and the state may be occuppied or empty ($d_N^{\dagger} d_N=1,0$, respectively) with no energy
cost. These two states are therefore degenerate in energy.
Solving the Bogoliubov-de Gennes (BdG) equations of the Kitaev 
Hamiltonian using open boundary conditions leads
to two zero energy modes that at the special point are perfectly localized at the edges
of the chain as $\delta$-function peaks (with exponential accuracy).

The phase diagram of the Kitaev model has three types of phases: two topological
phases in which there are gapless edge modes if the system is finite and two
trivial phases with no edge modes. In the various phases the system is gapped
and at the transition lines the gap closes, allowing the possibility of a change
of topology. The transition lines are located at $\Delta=0$ and at $|\mu|=2t$.
In terms of the spin model the transition line $\Delta=0$ is just the isotropic
$XY$ model which is known to be gapless. Positive values of $\Delta$ imply that
the exchange interaction is preferred along the $X$ direction and if $\Delta<0$
the preferred direction is along $Y$. The three phases of the Kitaev model have
a direct correspondance with the phases in the spin model: the topological phases
have as duals two phases with ordering along the $X$ or $Y$ directions and the
trivial phase has as dual a trivial paramagnetic phase in the spin model.
There is therefore a duality between the topological properties in the fermionic
Kitaev model and Landau-like ordering in the dual spin model as a result of
the exact Jordan-Wigner non-canonical non-local transformation \cite{cobanera,greiter}.

The structure of the phases can be understood in terms of the Majorana representation
of the fermionic operators and two phases are illustrated in Fig. \ref{fig1}.
The structure is particularly clear at these special points but due to the topologically
protected nature of the Hamiltonian (BDI class) their nature is not changed
as long as the gap does not close.

\subsection{Multiband system:
Two-band Schockley model}

\begin{figure}
\includegraphics[width=0.75\columnwidth]{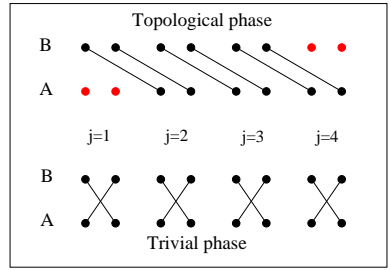}
\caption{\label{fig2}
(Color online) 
Phases of Schockley model. 
The symbols have the same meaning as in Fig. \ref{fig1}.
Trivial phase with $t_2=0$ and topological phase with $t_1=0$.
In the trivial phase the Majorana fermions are coupled to form fermionic
modes at the same site. In the topological phase the links are between
nearest-neighbor sites and there are four Majorana operators decoupled at the edges.
However, these give origin to zero-energy fermionic modes, one at each end of the
system.
}
\end{figure}

The Shockley model is a model of a dimerized system of spinless fermions with alternating
nearest-neighbor hoppings,  given by the Hamiltonian (see for instance \cite{yakovenko})
\be
H=\sum_{j=1}^N \psi^{\dagger}(j) \left[ U \psi(j) + V \psi(j-1) + V^{\dagger} \psi(j+1) 
\right)
\ee
where the $2 \times 2$ matrices $U$ and $V$ are given by
\be
U= \left(\begin{array}{cc}
0 & t_1^* \\
t_1 & 0  \end{array}\right); 
V= \left(\begin{array}{cc}
0 & t_2^* \\
0 & 0  \end{array}\right)
\ee
and the spinor $\psi$ represents two orbitals that are hybridized
by the matrices $U$ and $V$
\be
\psi(j) =
\left(\begin{array}{c}
\psi_a(j) \\
\psi_b(j) \end{array}\right).
\ee

We may as well define Majorana operators as
\bea
c_{j,A} &=& \frac{1}{2} \left( \gamma_{j,A,1} + i \gamma_{j,A,2} \right)\nonumber \\
c_{j,B} &=& \frac{1}{2} \left( \gamma_{j,B,1} + i \gamma_{j,B,2} \right)
\eea
Taking $t_1^*=t_1, t_2^*=t_2$, the Hamiltonian may be written as
\bea
H&=&\frac{i t_1}{2} \sum_{j=1}^N \left( \gamma_{j,A,1} \gamma_{j,B,2} +
\gamma_{j,B,1} \gamma_{j,A,2} \right)  \nonumber \\
&+&\frac{t_2}{4} \sum_{j=2}^N \left( \gamma_{j,A,1} \gamma_{j-1,B,1} +
\gamma_{j,A,2} \gamma_{j-1,B,2}  \right) \nonumber \\
&+& \frac{i t_2}{4} \sum_{j=2}^N \left( \gamma_{j,A,1} \gamma_{j-1,B,2}- 
i \gamma_{j,A,2} \gamma_{j-1,B,1}  \right) \nonumber \\
&+&\frac{t_2}{4} \sum_{j=1}^{N-1} \left( \gamma_{j,B,1} \gamma_{j+1,A,1} +
\gamma_{j,B,2} \gamma_{j+1,A,2} \right) \nonumber \\ 
&+& \frac{i t_2}{4} \sum_{j=1}^{N-1} \left(  \gamma_{j,B,1} \gamma_{j+1,A,2}- 
i \gamma_{j,B,2} \gamma_{j+1,A,1}  \right)
\eea

Taking $t_1=0$ we find that the Majorana fermions $\gamma_{1,A,1}, \gamma_{1,A,2},
\gamma_{N,B,1}, \gamma_{N,B,2}$ do not contribute and are zero energy modes.

In Fig. \ref{fig2} the structure of the Hamiltonian terms is presented
for two points in parameter phase that correspond to the trivial and the
topological phases. In the topological phase there are decoupled zero-energy
modes that are however fermionic in nature since the decoupled Majoranas
are located at the two end sites, $A$ and $B$, respectively.

\subsection{Multiband system: SSH model with triplet pairing}

\begin{figure}
\includegraphics[width=0.75\columnwidth]{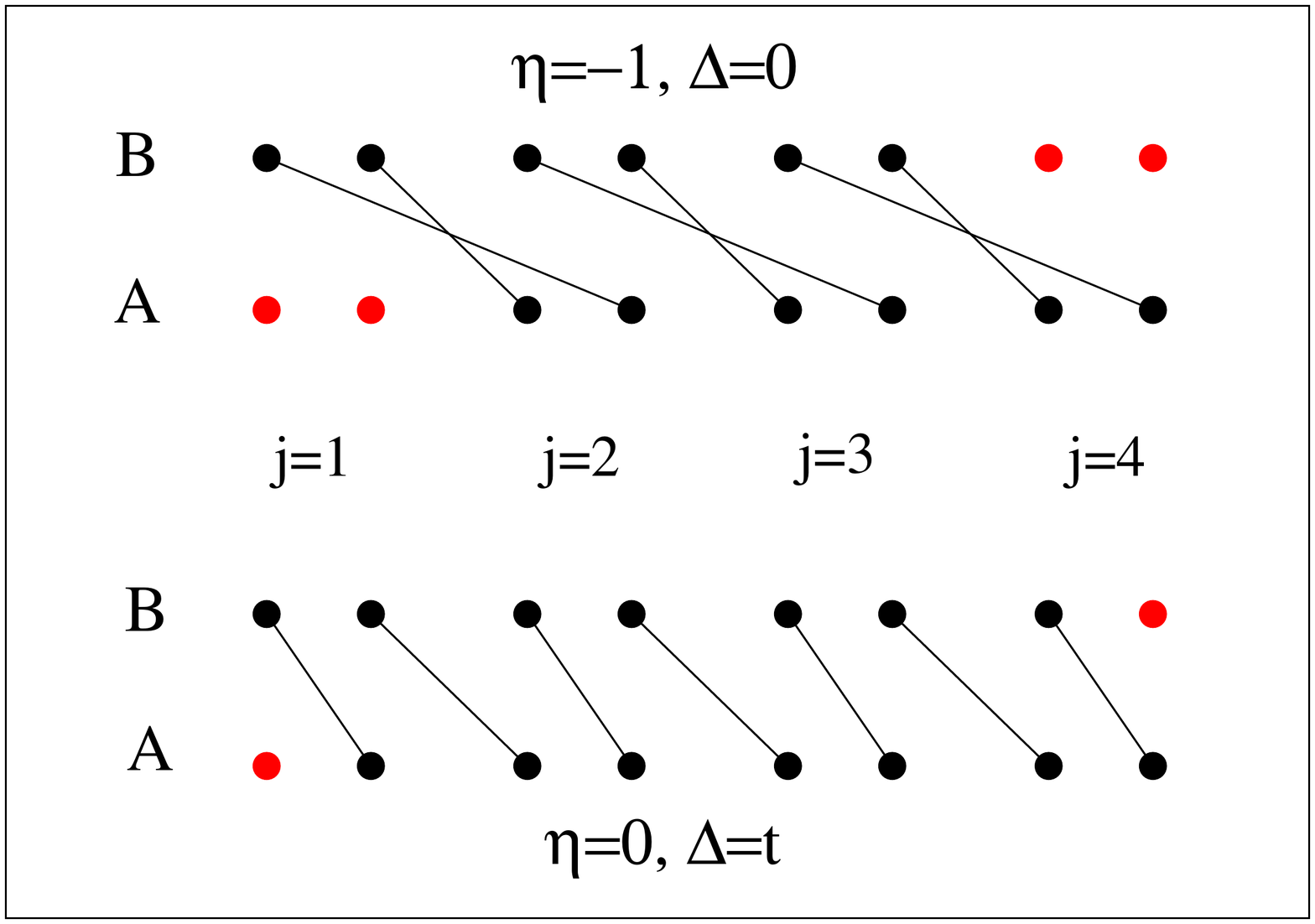}
\caption{\label{fig3}
(Color online) 
Phases of SSH-Kitaev model. The symbols have the same meaning as in Fig. \ref{fig1}.
When $\Delta=0$ the model reduces to the SSH model and for negative
$\eta$ the model is topologically non-trivial with edge states represented
by the decoupled Majorana operators. As in the Schockley model since at each
end site there are two decoupled Majoranas, these combine to form edge fermionic
modes. This constitutes phase SSH2 with $\eta=-1, \Delta=0$ and two edge modes.
If superconductivity is present, and there is no dimerization $\eta=0$, the model
reduces to the Kitaev model. The phase K1 with $\eta=0, \Delta=t$ has two
decoupled Majorana operators, one at each end, and therefore there is one Majorana
mode at each edge. The model interpolates between Majorana modes and fermionic
modes as the parameters change. There is also a trivial phase with no zero energy modes
denoted SSH0 which is similar to the trivial phase of the Schockley model.
}
\end{figure}

This model may be viewed as a dimerized Kitaev superconductor \cite{tanaka}.
The dimerization is parametrized by $\eta$ and the superconductivity
by $\Delta$.

This model is given by the Hamiltonian
\bea
H = -\mu & \sum_j & \left(  c_{j,A}^{\dagger} c_{j,A} + c_{j,B}^{\dagger} c_{j,B} \right)
\nonumber \\
-t & \sum_j &  \left[  (1+\eta) c_{j,B}^{\dagger} c_{j,A} + (1+\eta) c_{j,A}^{\dagger} c_{j,B} \right.
\nonumber \\
&+& \left. (1-\eta) c_{j+1,A}^{\dagger} c_{j,B} +(1-\eta) c_{j,B}^{\dagger} c_{j+1,A} \right]
\nonumber \\
+\Delta  & \sum_j &  \left[  (1+\eta) c_{j,B}^{\dagger} c_{j,A}^{\dagger} + (1+\eta) c_{j,A} c_{j,B} \right.
\nonumber \\
&+& \left. (1-\eta) c_{j+1,A}^{\dagger} c_{j,B}^{\dagger} +(1-\eta) c_{j,B} c_{j+1,A} \right]
\nonumber \\
& &
\eea
The model with no superconductivity ($\Delta=0$) is related to the Schockley model
taking $t_1=t(1+\eta)$ and $t_2=t(1-\eta)$. The region of $\eta>0$ corresponds to $t_1>t_2$
and vice-versa for $\eta<0$.
The Hamiltonian in real space mixes nearest-neighbor sites and also has
local terms. The local terms can be grouped in the matrix
\be
H_{j,j} = \left(\begin{array}{cccc}
-\mu & -t(1+\eta) & 0 & -\Delta (1+\eta) \\
-t(1+\eta) & -\mu  & \Delta (1+\eta) & 0  \\
0 & \Delta (1+\eta) & \mu & t(1+\eta) \\
-\Delta (1+\eta) & 0 & t(1+\eta) & \mu  \\
\end{array}\right)
\ee
The non-local terms to the nearest-neighbors can be written as
\be
H_{j,j+1} = \left(\begin{array}{cccc}
0 & 0 & 0 & 0 \\
-t(1-\eta) & 0  & -\Delta (1-\eta) & 0  \\
0 & 0 & 0 & 0 \\
\Delta (1-\eta) & 0 & t(1-\eta) & 0  \\
\end{array}\right)
\ee
and
\be
H_{j,j-1} = \left(\begin{array}{cccc}
0 & -t(1-\eta) & 0 & \Delta (1-\eta) \\
0 & 0  & 0 & 0  \\
0 & -\Delta (1-\eta) & 0 & t(1-\eta) \\
0 & 0 & 0 & 0  \\
\end{array}\right)
\ee

\begin{figure}
\includegraphics[width=0.75\columnwidth]{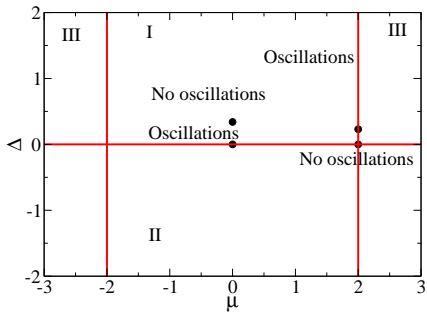}
\caption{\label{fig4}
(Color online) 
Phase diagram of Kitaev model. Regions I and II are topologically non-trivial and
region III is trivial. The points separate regions where critical quenches lead or not
to oscillations in the survival probability of a Majorana state of region I.
}
\end{figure}

In momentum space this model is given by an Hamiltonian matrix of the form
\be
\label{hmm}
H_k = \left(\begin{array}{cccc}
-\mu & z(k) & 0 & w(k) \\
z^*(k) & -\mu  & -w^*(k) & 0  \\
0 & -w(k) & \mu & -z(k) \\
w^*(k) & 0 & -z^*(k) & \mu  \\
\end{array}\right)
\ee
where this matrix acts on the spinors
\be
\left(\begin{array}{c}
c_A(k) \\
c_B(k) \\
c_{A}^{\dagger}(-k) \\
c_{B}^{\dagger}(-k) \\
 \end{array}\right)
\ee
and
\bea
z(k) &=& -t \left[(1+\eta)+(1-\eta) e^{-ik} \right] \nonumber \\
w(k) &=& -\Delta \left[(1+\eta)-(1-\eta) e^{-ik} \right] 
\eea

In terms of Majorana operators the Hamiltonian is written as
\bea
H &=& -\frac{\mu}{2} \sum_{j=1}^N \left( 2+i\gamma_{j,A,1} \gamma_{j,A,2}
+ i \gamma_{j,B,1} \gamma_{j,B,2} \right) \nonumber \\
&-& \frac{it}{2} (1+\eta) \sum_{j=1}^N \left(
\gamma_{j,B,1} \gamma_{j,A,2} + \gamma_{j,A,1} \gamma_{j,B,2} \right) \nonumber \\
&-& \frac{it}{2} (1-\eta) \sum_{j=1}^{N-1} \left(
\gamma_{j+1,A,1} \gamma_{j,B,2} + \gamma_{j,B,1} \gamma_{j+1,A,2} \right) \nonumber \\
&+& \frac{i\Delta}{2} (1+\eta) \sum_{j=1}^{N} \left(
\gamma_{j,A,1} \gamma_{j,B,2} + \gamma_{j,A,2} \gamma_{j,B,1} \right) \nonumber \\
&+& \frac{i\Delta}{2} (1-\eta) \sum_{j=1}^{N-1} \left(
\gamma_{j,B,1} \gamma_{j+1,A,2} + \gamma_{j,B,2} \gamma_{j+1,A,1} \right)
\nonumber \\
& &
\eea

\begin{figure*}
\includegraphics[width=0.3\textwidth]{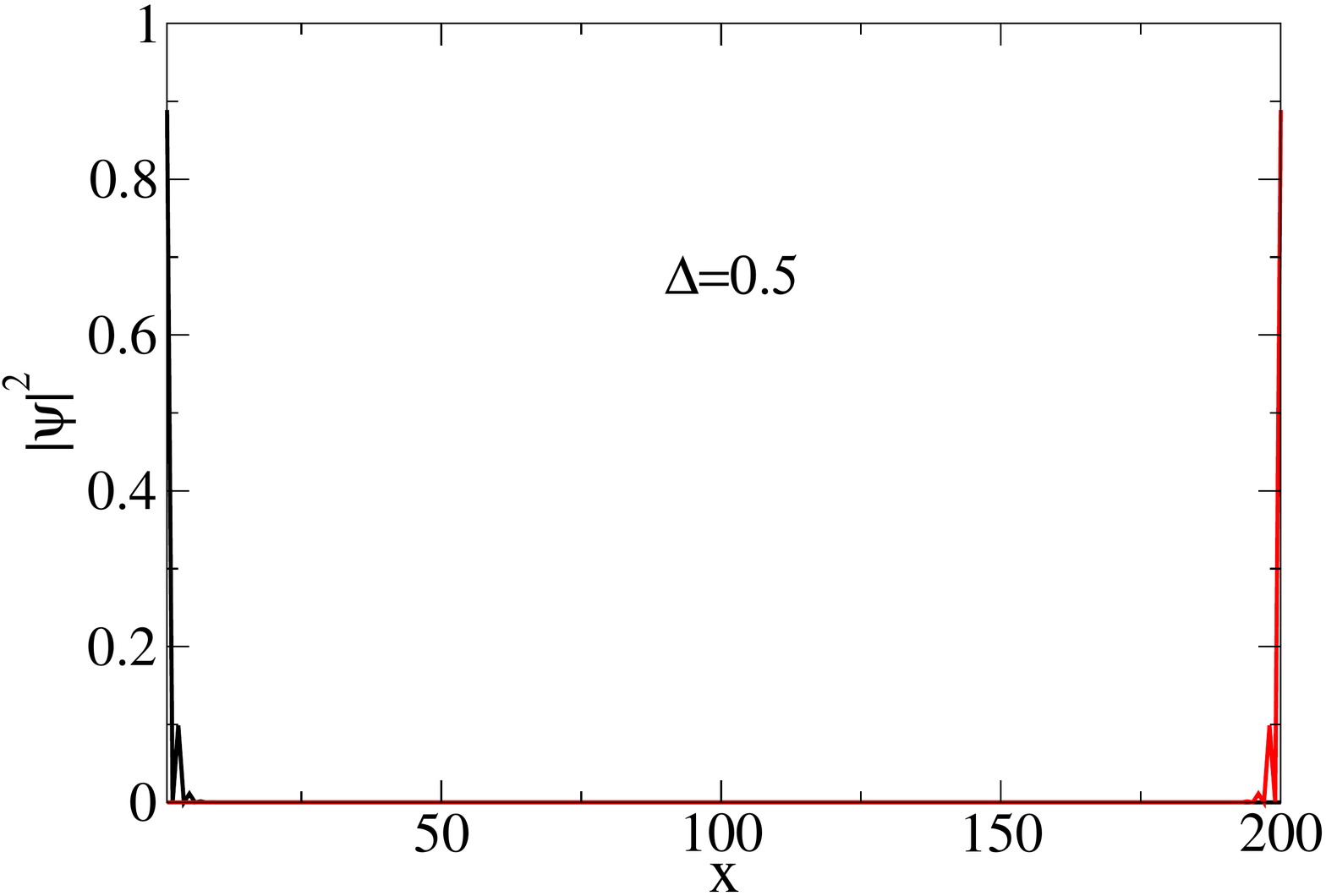}
\includegraphics[width=0.3\textwidth]{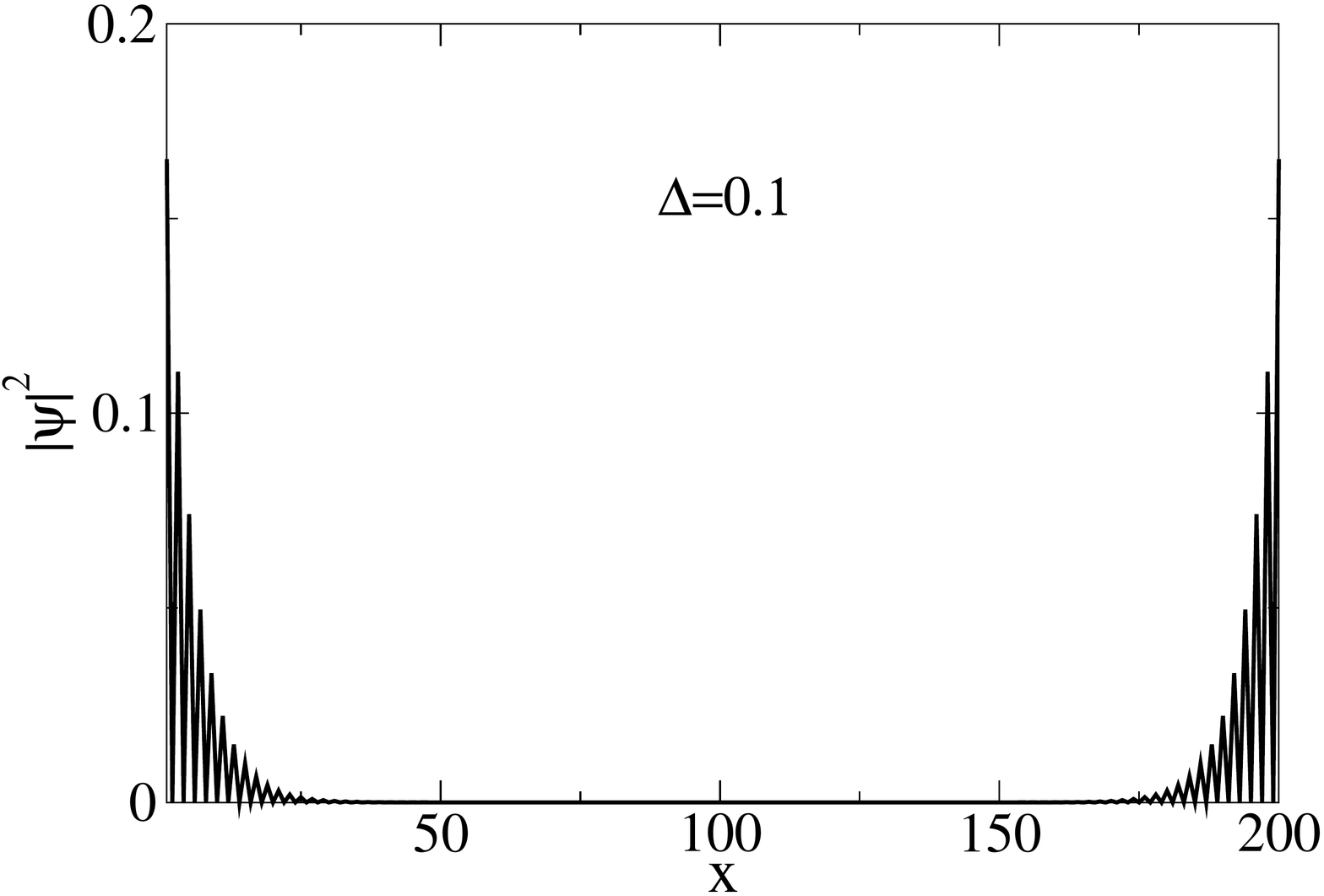}
\includegraphics[width=0.3\textwidth]{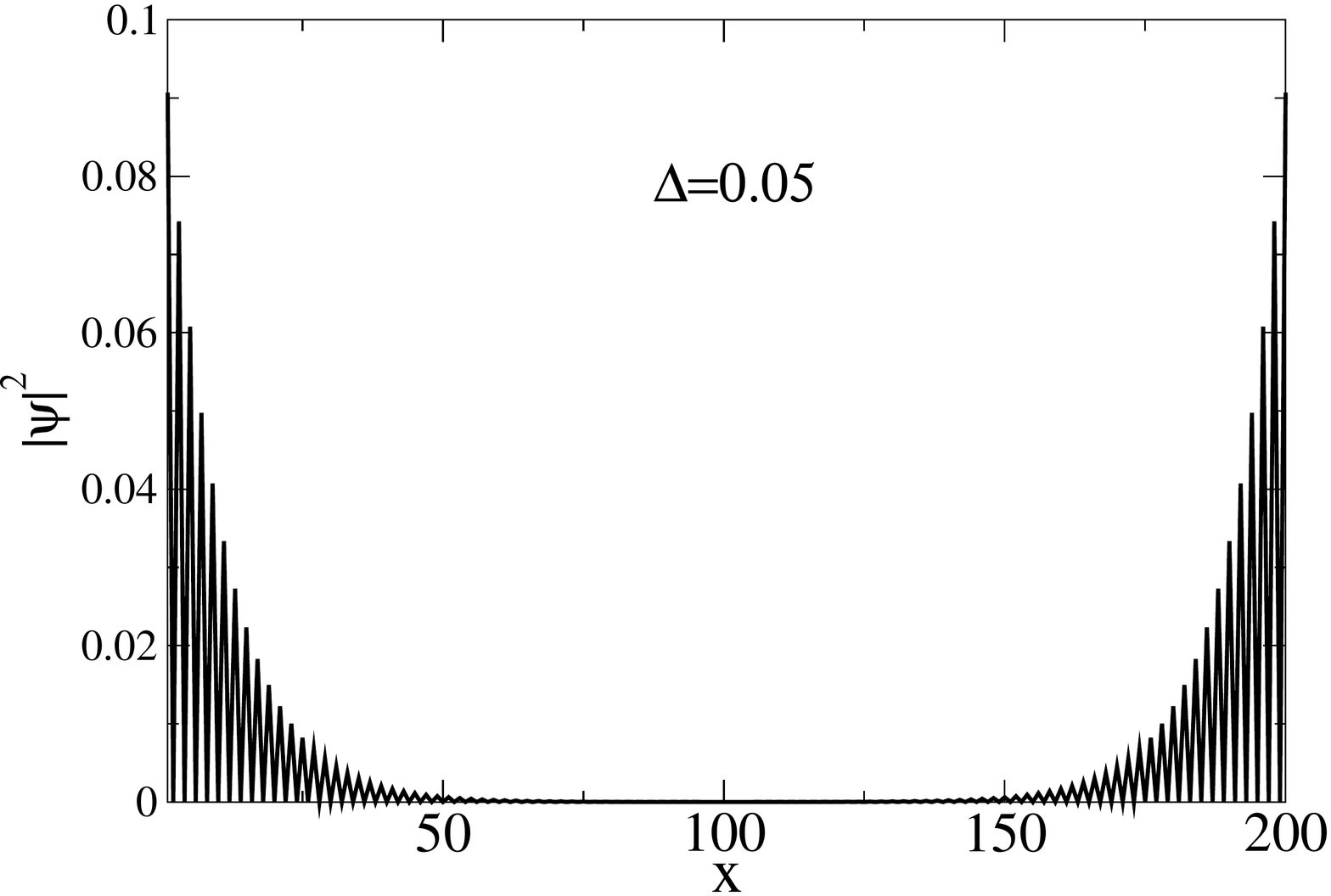}
\includegraphics[width=0.3\textwidth]{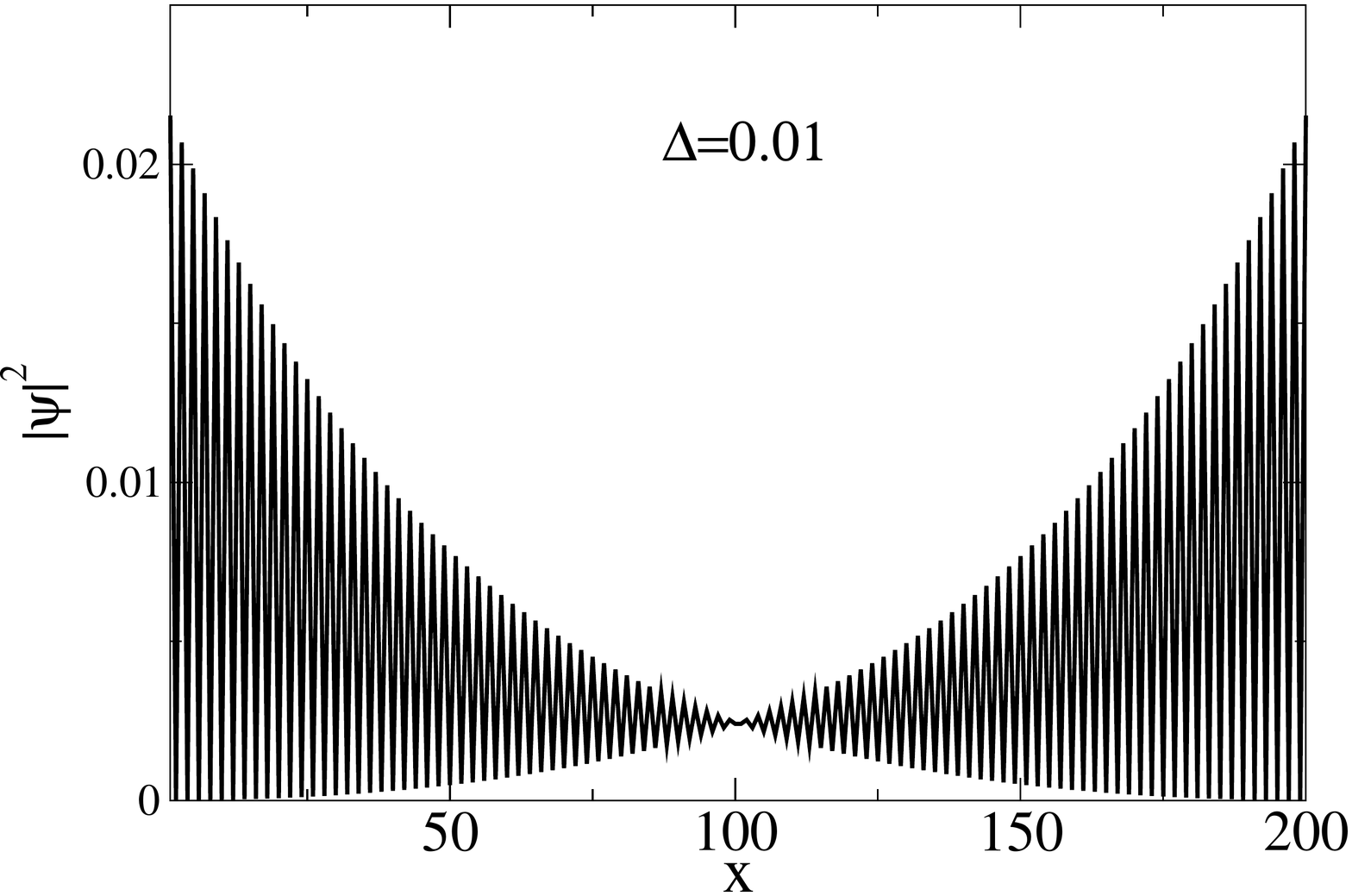}
\includegraphics[width=0.3\textwidth]{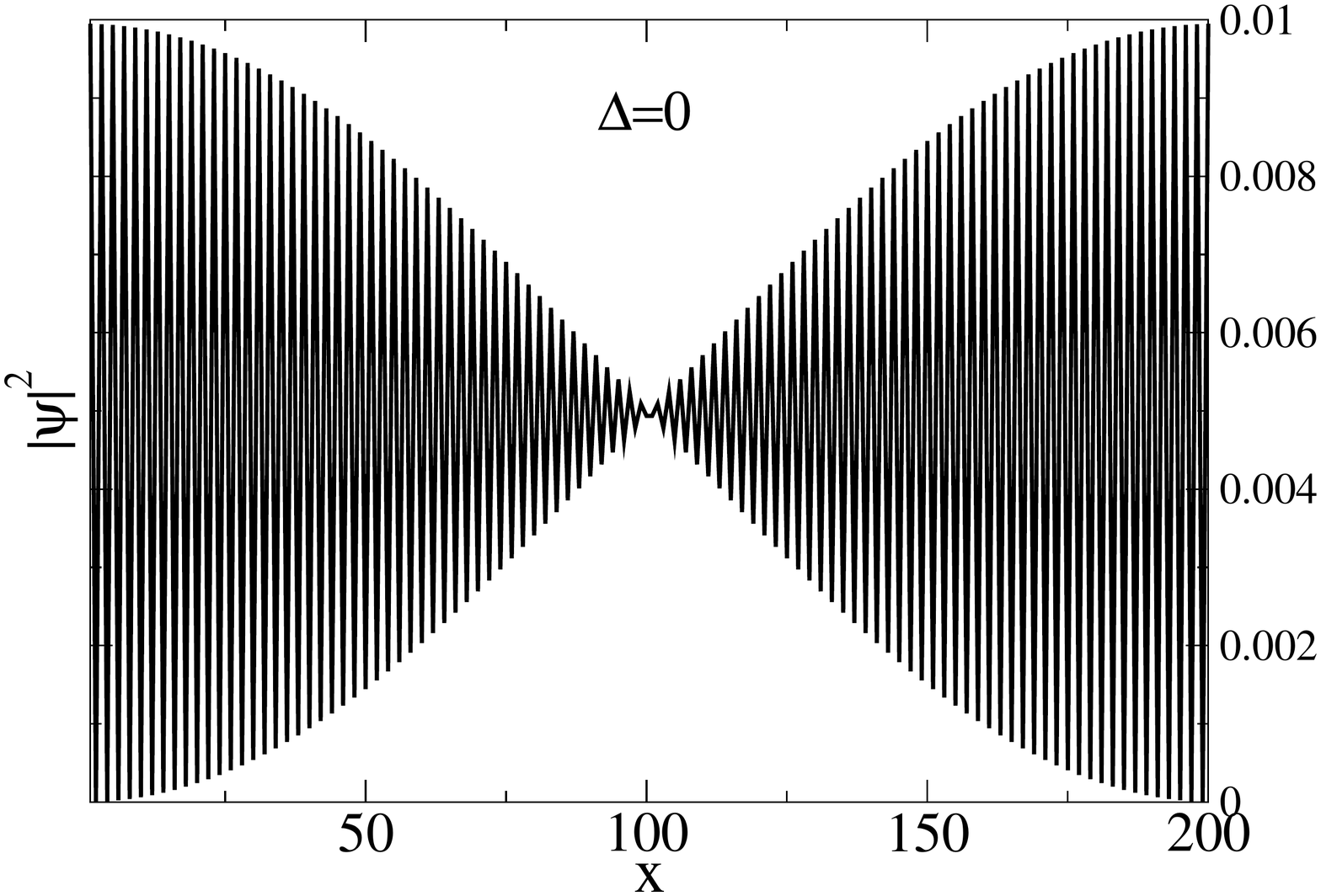}
\includegraphics[width=0.3\textwidth]{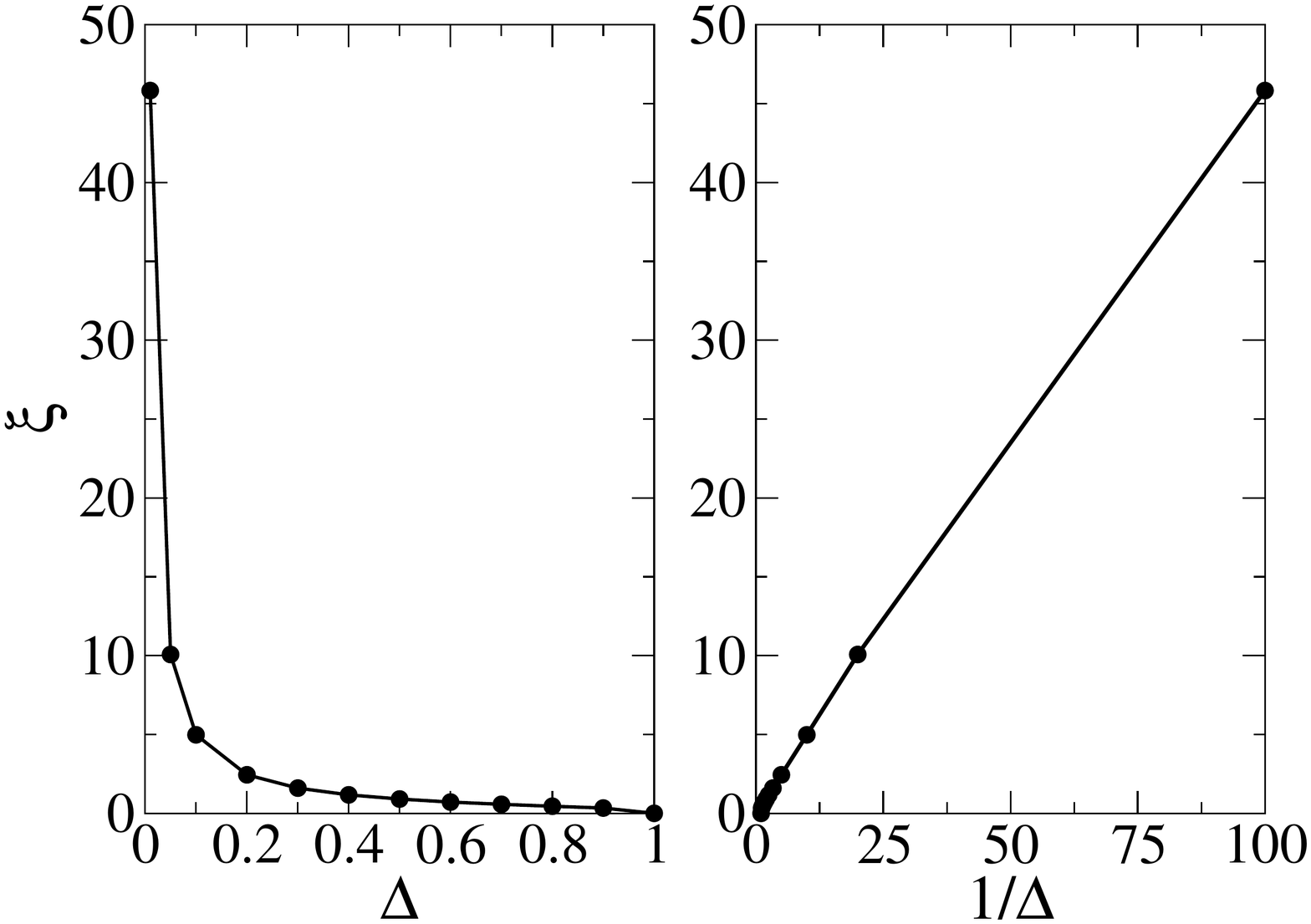}
\caption{\label{fig5}
(Color online) 
Zero energy mode wave functions of Kitaev model for different points approaching
the critical point at $\mu=0$. In the last panel we show the dependence of the decay length
of the wave functions as a function of $\Delta$ and $1/\Delta$. The results are for a system size of $N=200$. 
At the point $\Delta=1$ the wave function is strictly local at one edge of the system.
Each Majorana state is perfectly localized with a decay length $\xi=0$. 
For $\Delta=0.5$ the decay length is also very small and the two Majorana modes
(in black, left edge, and in red, right edge) are decoupled.
For smaller values of $\Delta$ the two Majorana modes are coupled and each is peakd at both ends
of the chain. For very small values of $\Delta$ the modes get extended as one tends to the
gapless regime at $\Delta=0$.
}
\end{figure*}

Taking as before $\mu=0$ we have a couple of special points: i) $\eta=-1$ and
$\Delta=0$ we have a state similar to the SSH or Schockley models with two
fermionic-like zero energy edge states, since the four operators
$\gamma_{1,A,1}, \gamma_{1,A,2}; \gamma_{N,B,1}, \gamma_{N,B,2}$ are missing from
the Hamiltonian.
ii) $\eta=0$ and $t=\Delta$ is a Kitaev like state since there are two Majorana operators
missing from the Hamiltonian, $\gamma_{1,A,1}$ and $\gamma_{N,B,2}$, one from each end.
iii) An example of a trivial phase is the point $\eta=1$ and $\Delta=0$ in which case there
are no zero energy edge states.
In Fig. \ref{fig3} the phases with edge modes are presented for special points
in parameter space. 
This model provides a testing ground for the comparison of fermionic and Majorana
edge modes. Also, in some regimes it displays finite energy modes that are localized
at the edges of the chain, as obtained before in other multiband models \cite{rio}.

In addition to direct measurements of tunneling density of states of Majorana edge
states, measurements of the differential conductance at the interface between a lead
and a topological superconductor have been proposed as a way to detect Majorana modes.
In particular, with a metallic lead one expects a zeo-bias peak in the differential
conductance, if zero-energy modes are present in the superconducting side.
In the presence of Majorana modes one expects a vanishing conductance if the number of
Majorana modes is even and a quantized value of $2e^2/h$, if the number of modes is odd \cite{law,wimmer}.
These may be due to edge chiral modes or, in a p-wave superconductor, associated with
a vortex \cite{law}. In the case of the dimerized SSH model here considered
it has been shown \cite{tanaka} that the fermionic edge modes do not contribute to the
conductance and, therefore, provides a method to distinguish the various phases, 
since in the regime that is Kitaev-like the conductance is quantized, as expected, while
in the SSH regimes it vanishes. However, other types of zero-energy states due for instance
to disorder or temperature effects, also give origin to zero-bias peaks in the conductance \cite{liu1}.
A possible way to clearly identify a zero mode as a Majorana has been proposed \cite{peng} selecting a
superconducting lead with the prediction of peaks at the value of the gap of the conventional
superconductor with a quantized conductance of the form $(4-\pi)2e^2/h$.
In this work we will analyse possible distinctive signatures of the edge modes 
in the dynamics following a sudden quantum quench.

\section{Dynamics of edge modes of Kitaev model}

\subsection{Single quench}

\begin{figure*}
\includegraphics[width=0.4\textwidth]{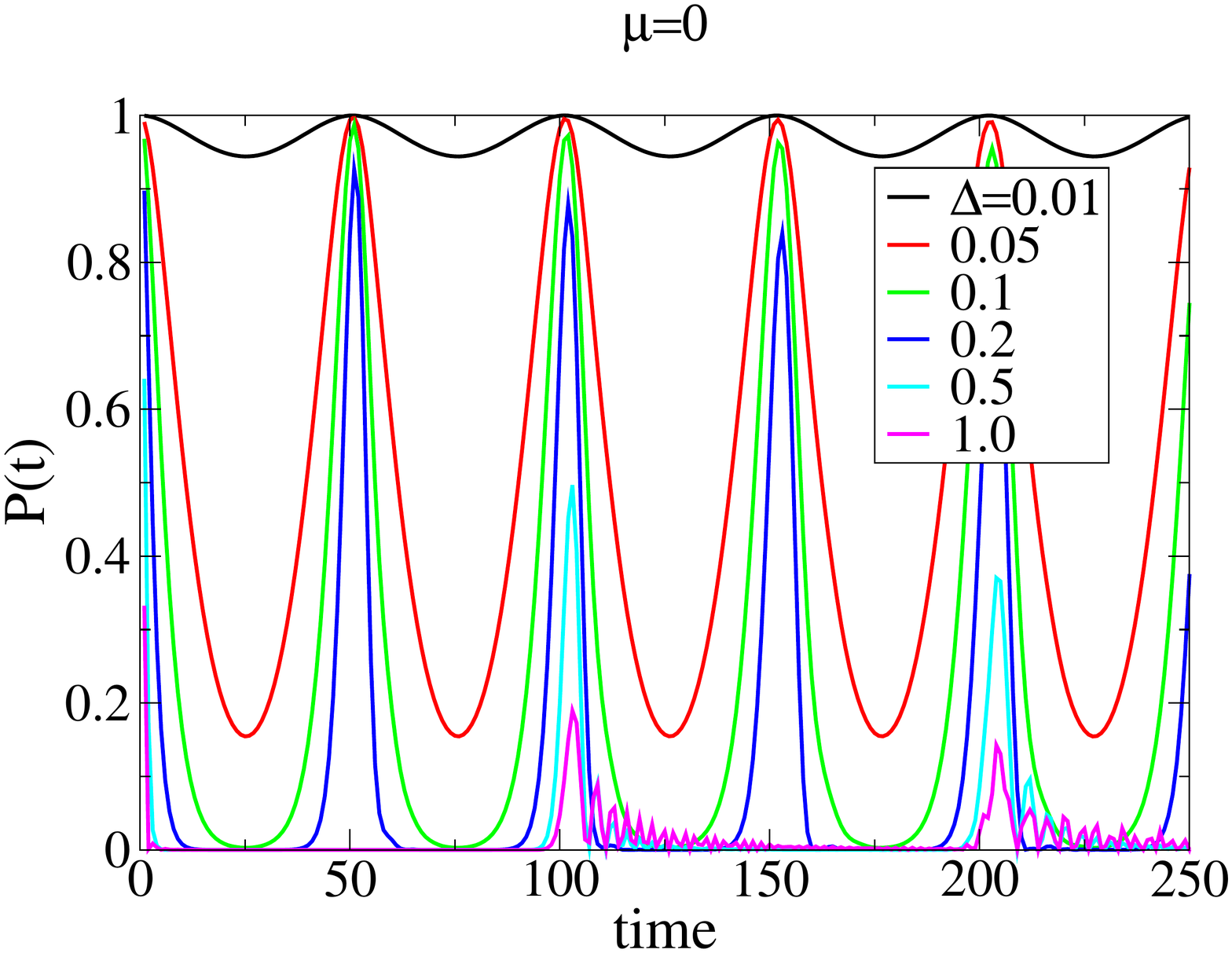}
\includegraphics[width=0.4\textwidth]{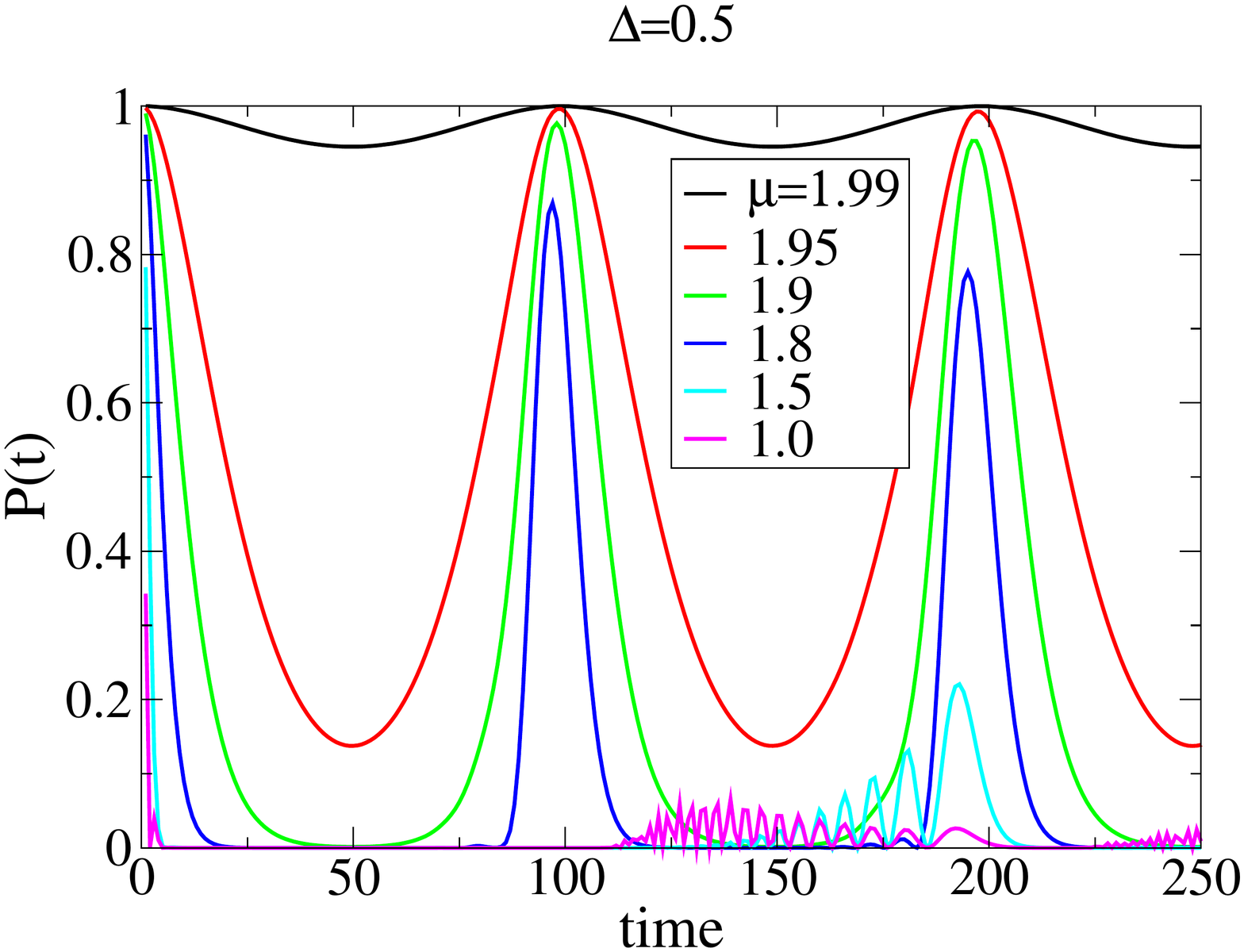}
\includegraphics[width=0.4\textwidth]{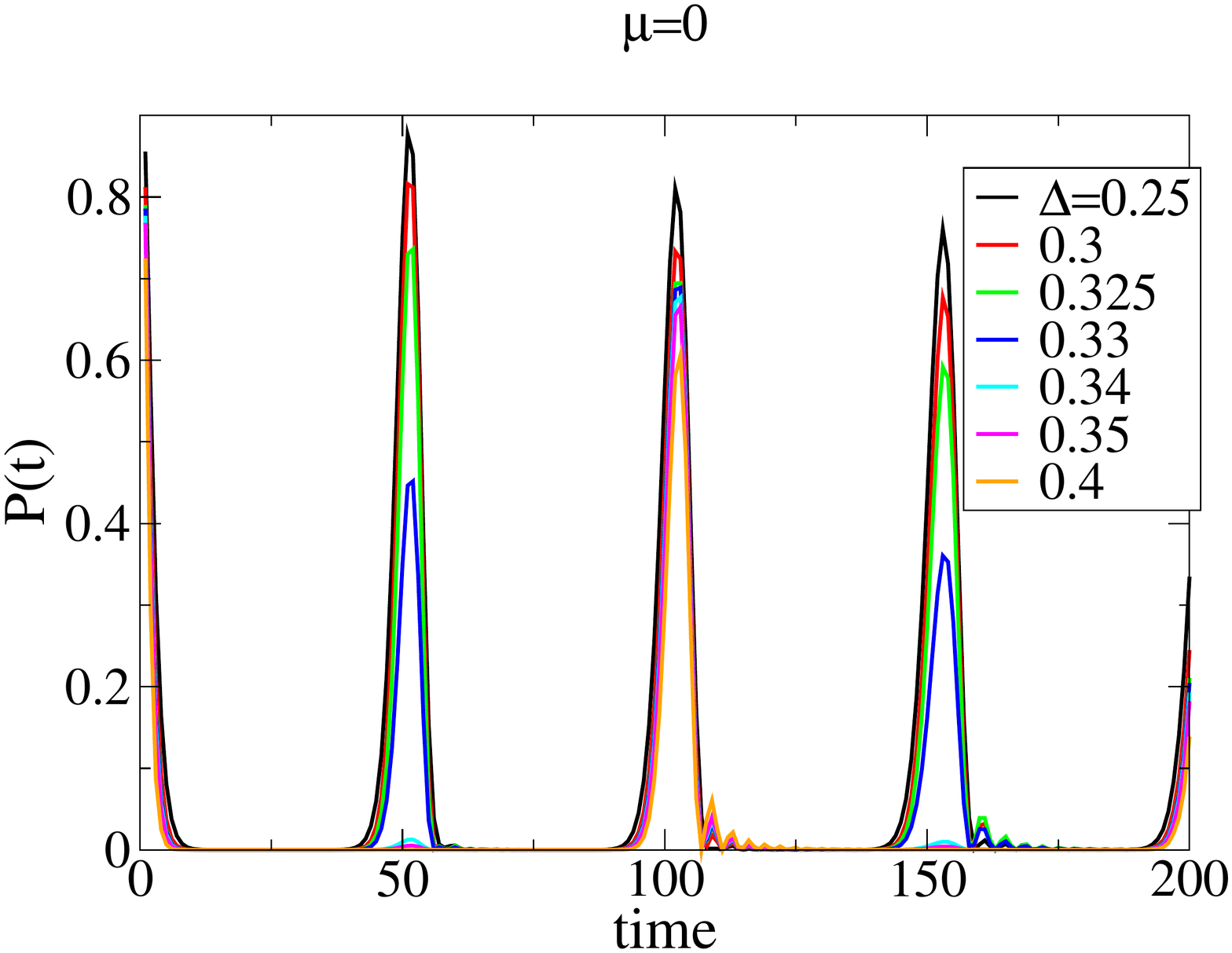}
\includegraphics[width=0.4\textwidth]{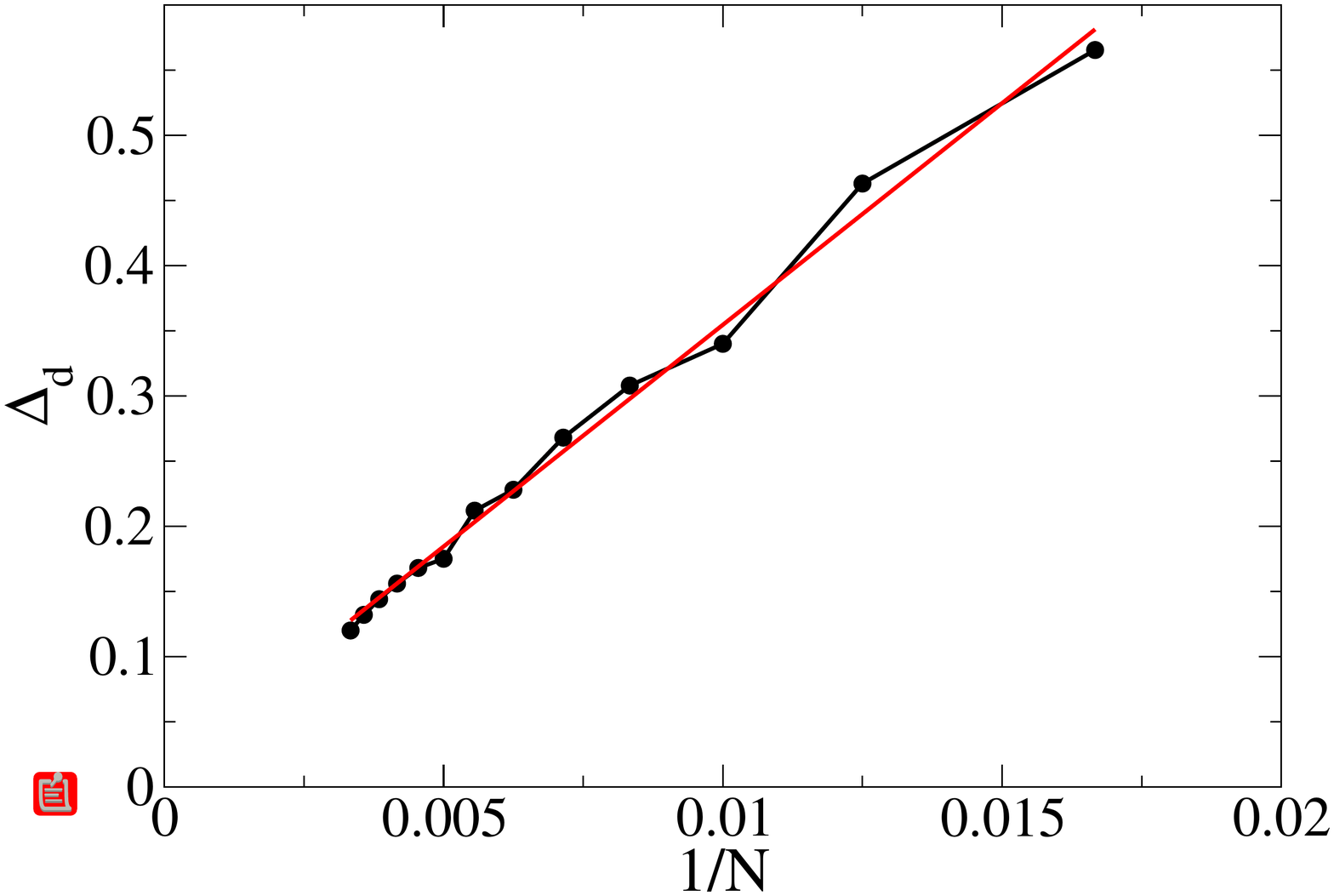}
\caption{\label{fig6}
(Color online) 
In the first two panels the survival probability, $P(t)$, of a Majorana mode in 
the Kitaev model is shown as one approaches critical points. In the third panel the
crossover to period doubling is shown as one approches the critical region. 
In the fourth panel the linear dependence of the point of crossover, $\Delta_d$, on $1/N$ is shown. 
}
\end{figure*}

In Fig. \ref{fig4} we show the phase diagram of the Kitaev model as a function of the chemical potential $\mu$
and $\Delta$. Regions $I$ and $II$ are topologically non-trivial and
regions $III$ are trivial, as discussed above. 
Consider first $\mu=0$ and quenches where one varies $\Delta$, or a fixed $\Delta$ and changing $\mu$.
The points separate the regions where one
finds or does not find oscillations in the survival probability of a Majorana state 
of a topologically
non-trivial phase after a quench
to a critical point, for a
given system size. 
In the case of $\mu=0$ the critical point is located at $\mu=0,\Delta=0$ and in the second case there
is a line of critical points at $\mu=2t$.
For instance, in the quench from the topological phase
$I$ to the critical point at $\mu=0, \Delta=0$, the point is located as
$N=100, \Delta=0.34$, $N=200, \Delta=0.18$, $N=400, 0.05<\Delta <0.1$.
The points at $\mu=2$
separate two regions for which making a quench from region $I$ to the
critical line $\mu=2$,
one may find oscillations, if the initial point
is not very far from the critical line.  
In the vicinity of the two critical lines of points (around $\mu=2t,\Delta=0$),
no matter how close the initial
point is to the critical line, one does not find oscillations.

In Fig. \ref{fig5} is shown the absolute value squared of the lowest energy
eigenvector for $N=200$ for diferent values of $\Delta$ and keeping $\mu=0$. 
Note that
if one is very close to the transition point the wave functions are not very localized.
At $\Delta=0$ the state is extended since the system is gapless. A similar behavior
is observed for small values of $\Delta$ and, as $\Delta$ increases, the decay length decreases
significantly. The decay length as a function of $\Delta$ is shown in the last panel.
The fit of the wave function dependence with distance, $x=j$, from the edge of the system,
is of an exponential form $|\psi|^2(2n+1)=\psi_0 e^{-x/\xi}$ (here $x=2n+1$ since $|\psi|^2$
oscillates), with $\xi$ the decay length.

In Fig. \ref{fig6}
the survival probability as a function of time for various critical quenches is presented.
In the first panel are shown the oscillations of $P(t)$ as one quenches from a given
value of $\Delta$ to the critical point $\mu=0, \Delta =0$ maintaining
$\mu=0$. 
For small deviations of the initial value of $\Delta$ from
the critical point, $P(t)$ is close to $1$ and as one increases the
distance from the critical point the amplitude decreases considerably.
The oscillations are quite smooth and clear until the amplitude has
decreased enough to reach zero. Beyond this point there is a periodicity
but no longer oscillations since there are increasing regions where $P(t)$
basically vanishes. In this case it seems more like the revival times
of non-critical quenches, even though the curves are still smooth. 
Beyond a given value of $\Delta_d$ there is a {\it period doubling}.
Also, after this period doubling the survival probability looses its regular
periodic behavior and shows more oscillations of smaller periods and amplitude
decays that are similar to results previously found in quenches away from
critical points \cite{rajak,pre}. 
In the second panel are shown quenches to the critical line $\mu=2$ keeping
$\Delta=0.5$ and decreasing the chemical potential. The behavior is similar
to the first panel.
In the third panel is shown in greater detail the {\it crossover} to period
doubling for the transition to the critical point. 
In the fourth panel we show the scaling of the point of crossover, $\Delta_d$,
when the period doubling
takes place. It scales linearly with $1/N$. 

\begin{figure}
\includegraphics[width=0.85\columnwidth]{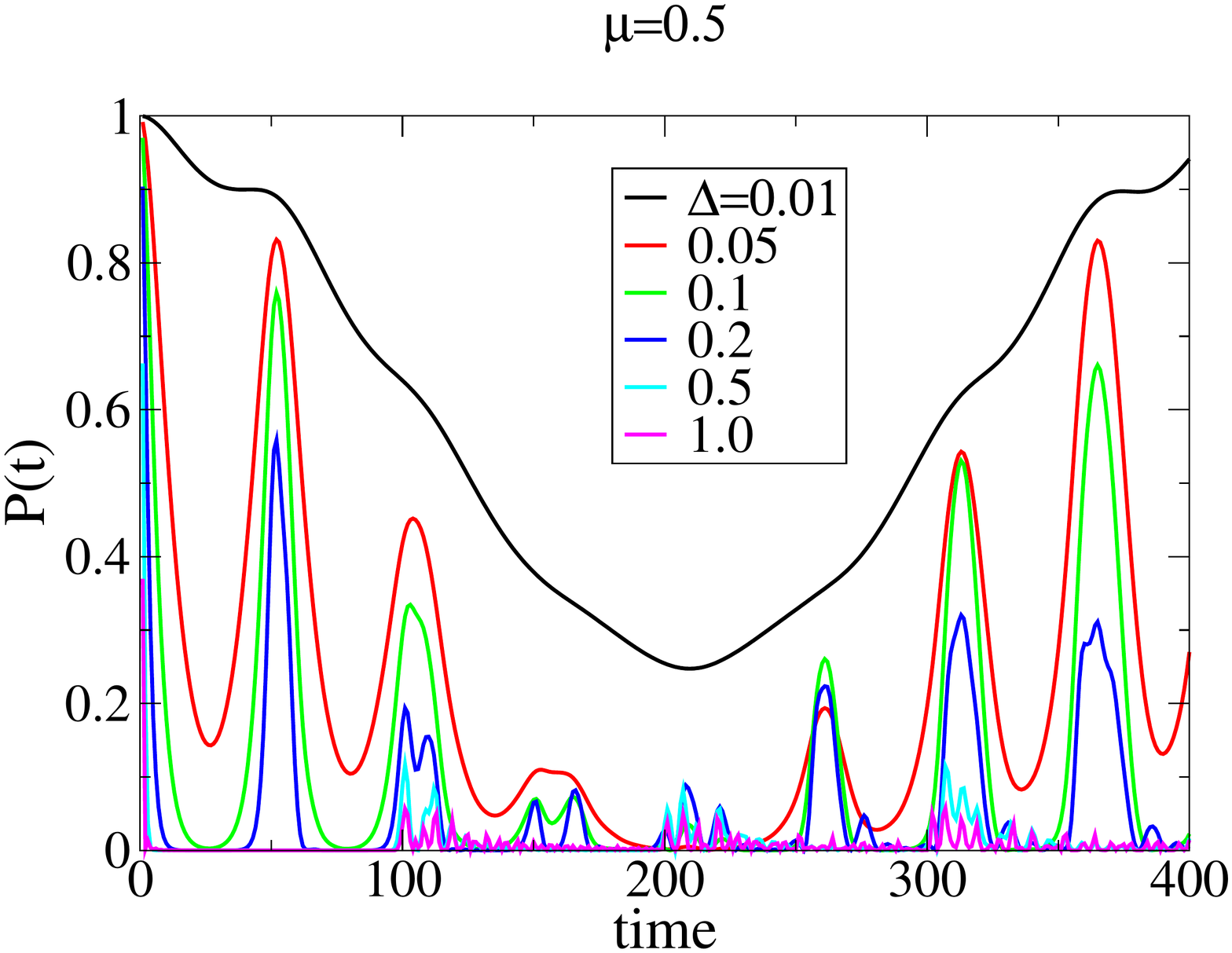}
\includegraphics[width=0.85\columnwidth]{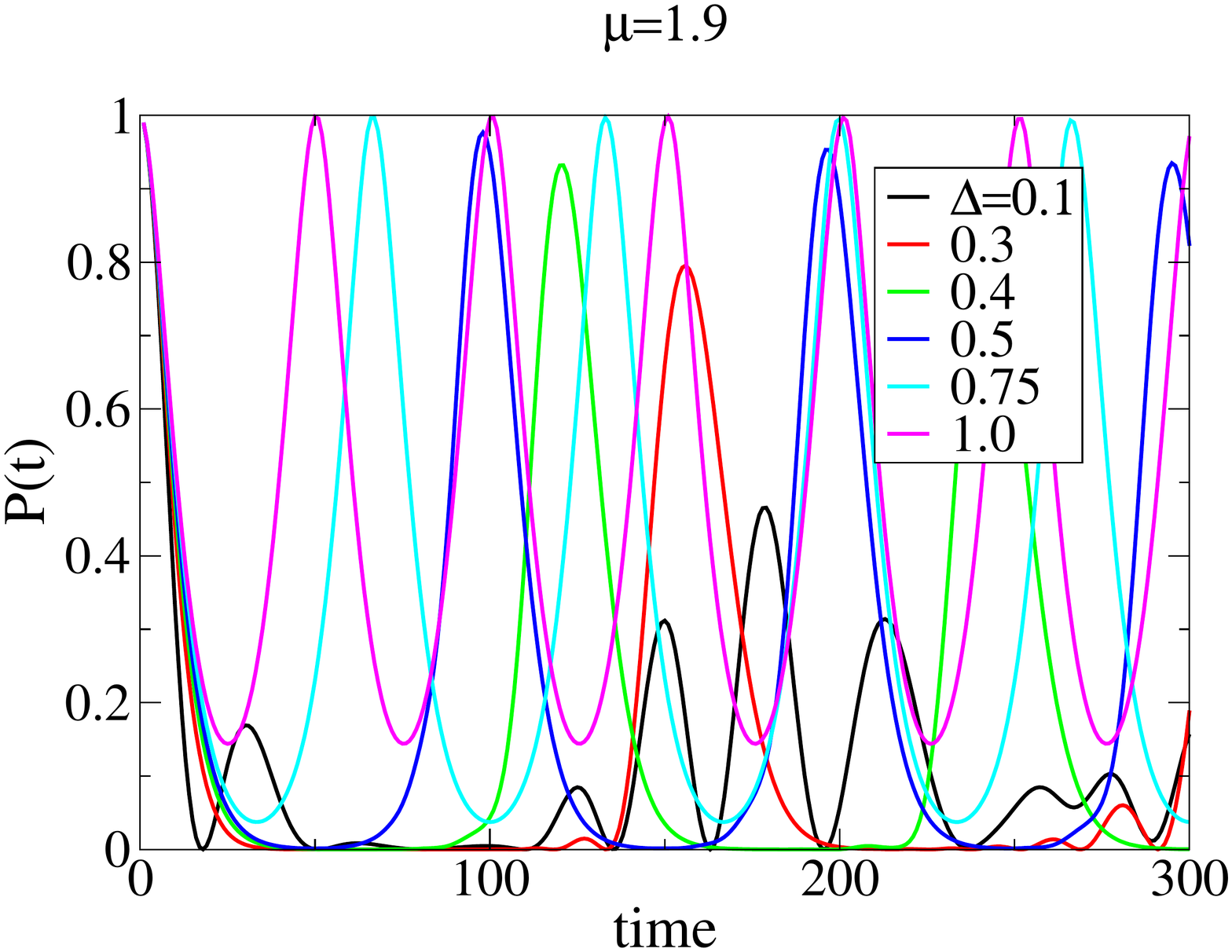}
\caption{\label{fig7}
(Color online) 
Left panel: at finite chemical potential the critical quench to the line $\Delta=0$ does not lead to oscillations
of the survival probability of the Majorana mode. Close to the intersection of the two critical lines near
$\mu=2t,\Delta=0$ there are no oscillations as well as shown in the right panel.
}
\end{figure}

In the first panel of Fig. \ref{fig7}
are shown quenches, keeping $\mu=0.5$, to the critical line
$\Delta=0$. In this case there are no oscillations.
In the second panel the quench to the critical line $\mu=2$ from the initial point
$\mu_i=1.9$ is shown for different values of $\Delta$. For small values of $\Delta$
there are no oscillations and as $\Delta$ increases the oscillations appear.
The results in this figure are for $N=100$.

\begin{figure}
\includegraphics[width=0.75\columnwidth]{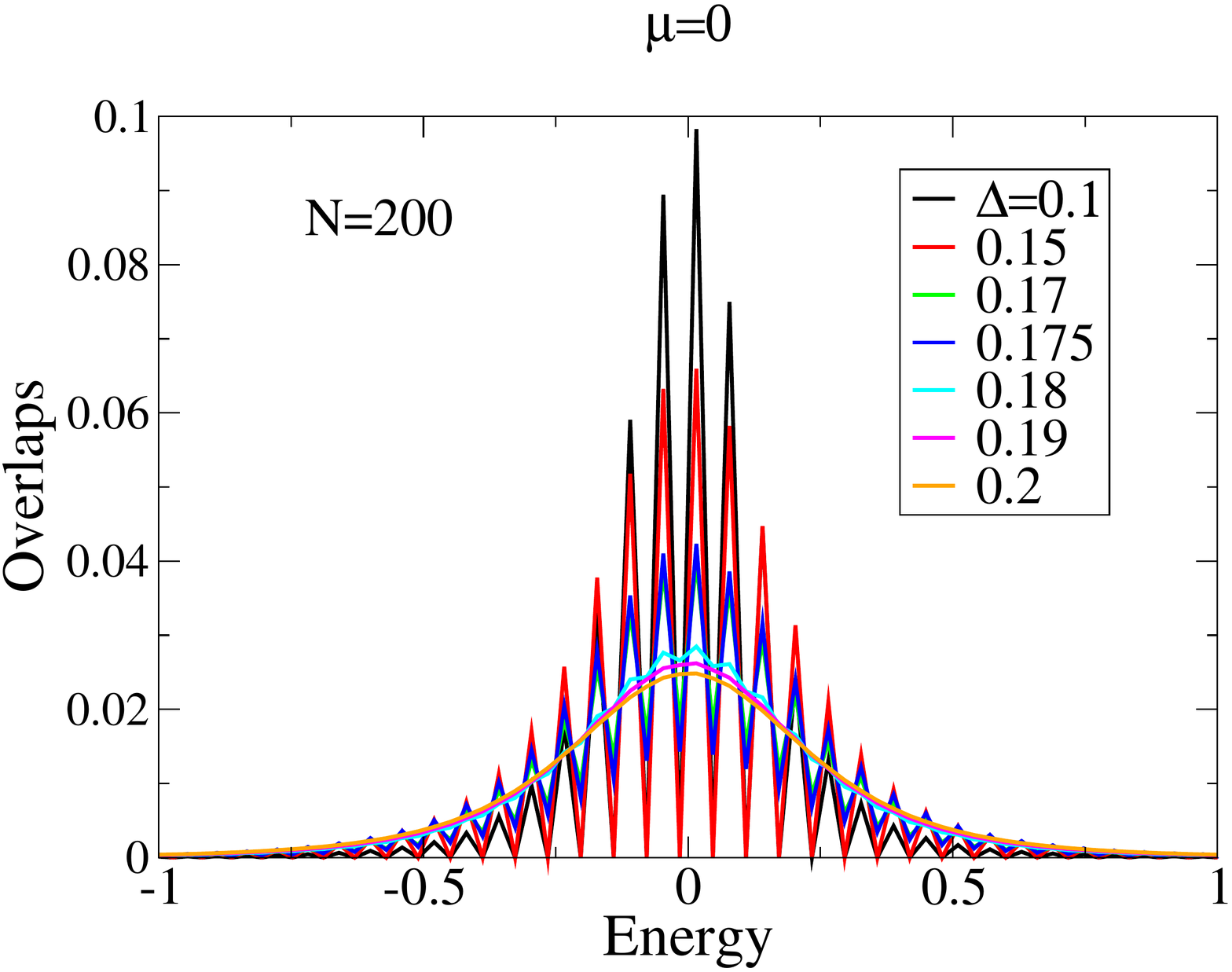}
\includegraphics[width=0.75\columnwidth]{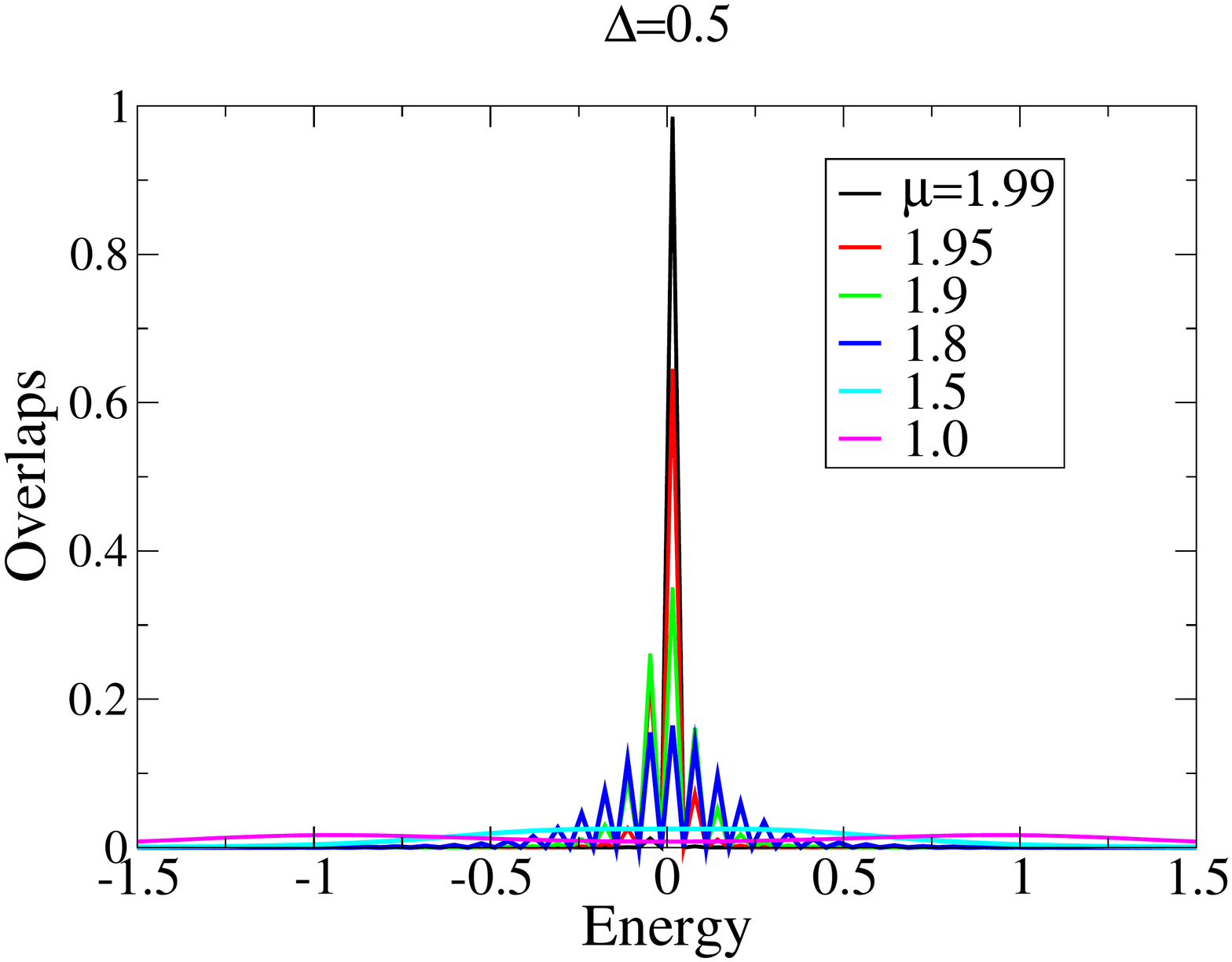}
\includegraphics[width=0.75\columnwidth]{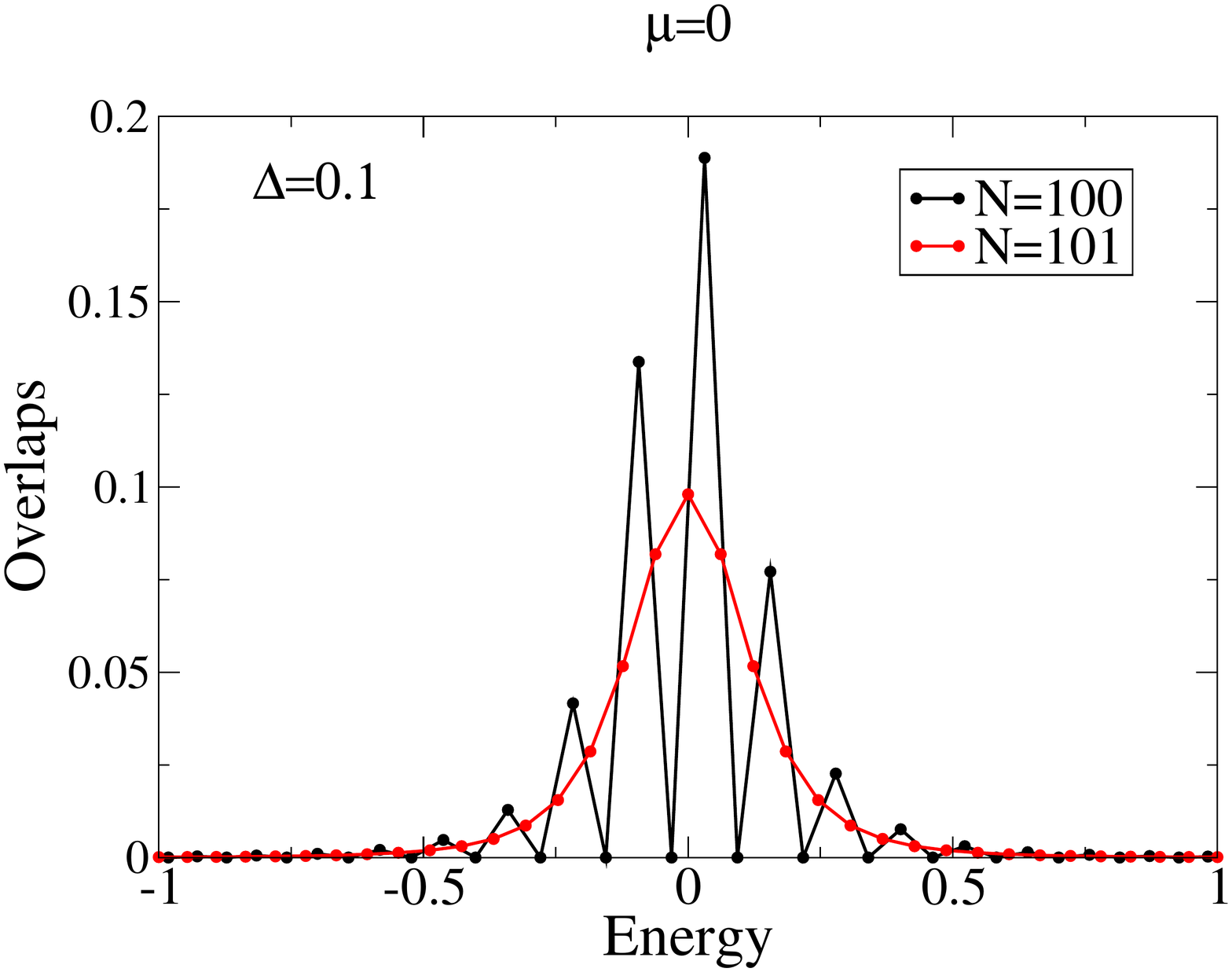}
\caption{\label{fig8}
(Color online) 
Overlaps for the Kitaev model as a function of energy. Top panel the
critical point is $\mu=0,\Delta=0$ and in the middle panel is
$\mu=2,\Delta=0.5$.
In the third panel the even-odd effect in the overlaps is shown.
}
\end{figure}

The survival probability is determined by the various energies of
the final Hamiltonian eigenstates and the overlaps to the initial state.
In Fig. \ref{fig8} the overlaps between the initial lowest
energy state (Majorana mode) and all the final state eigenvectors are shown,
as a function of their energies, for the cases of Fig. \ref{fig6}, for $N=200$. 
In general, the overlaps are peaked at the lowest energies. 
There is a clear separation of regimes as one reaches
the crossover region where the period doubling occurs. At small values
of $\Delta$ the overlaps oscillate between finite values and zero
values, as we move accross the energy eigenvalues. This is probably
a parity effect distinguishing even and odd number of sites. 
This is confirmed in the third panel where
the cases of $N=100$ and $N=101$ are compared. If $N=101$ the oscillations
in the overlaps are absent. 
It can be noted that the overlaps are very sharp around the lowest
energy states. As the crossover occurs the overlaps are no longer zero
at some energy eigenvalues and actually become very smooth. This means
that the contributions from the various energy states changes, the
time behavior is affected and the clean oscillations are no longer
observed. In order to have clean oscillations one needs contributions
from few energy levels. A perfect oscillation requires finite overlaps
to two states and the frequency of the oscillations is the difference
in their energy values. In general, the overlaps have very different
magnitudes to the two states and the period of oscillations shown in
$P(t)$ depends on their magnitudes. Adding significant contributions
from other energy eigenstates leads first to modulated oscillations
and then to a complicated time dependence.

It was argued before \cite{rajak} that the energy spectrum of the final
state of the quench is important to determine if there are oscillations.
In some cases there is indeed a regular spacing of the final state
energies and in others no. If the regular spacing is observed we should
find a dominant frequency but also many harmonics. However, the
role of the overlaps is more significant because it clearly selects which
energy states actually contribute \cite{haque}. 

\begin{figure}
\includegraphics[width=0.75\columnwidth]{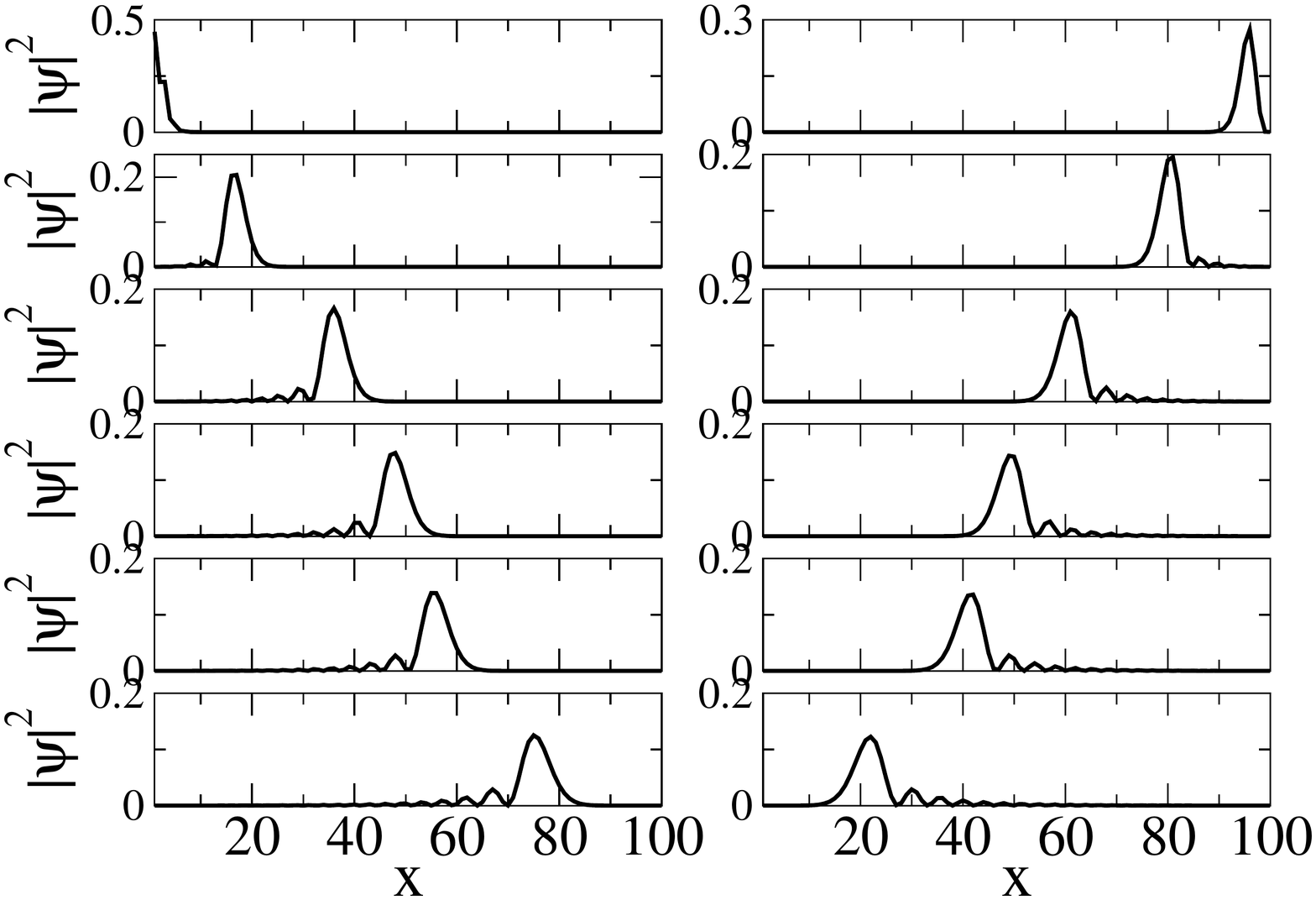}
\includegraphics[width=0.75\columnwidth]{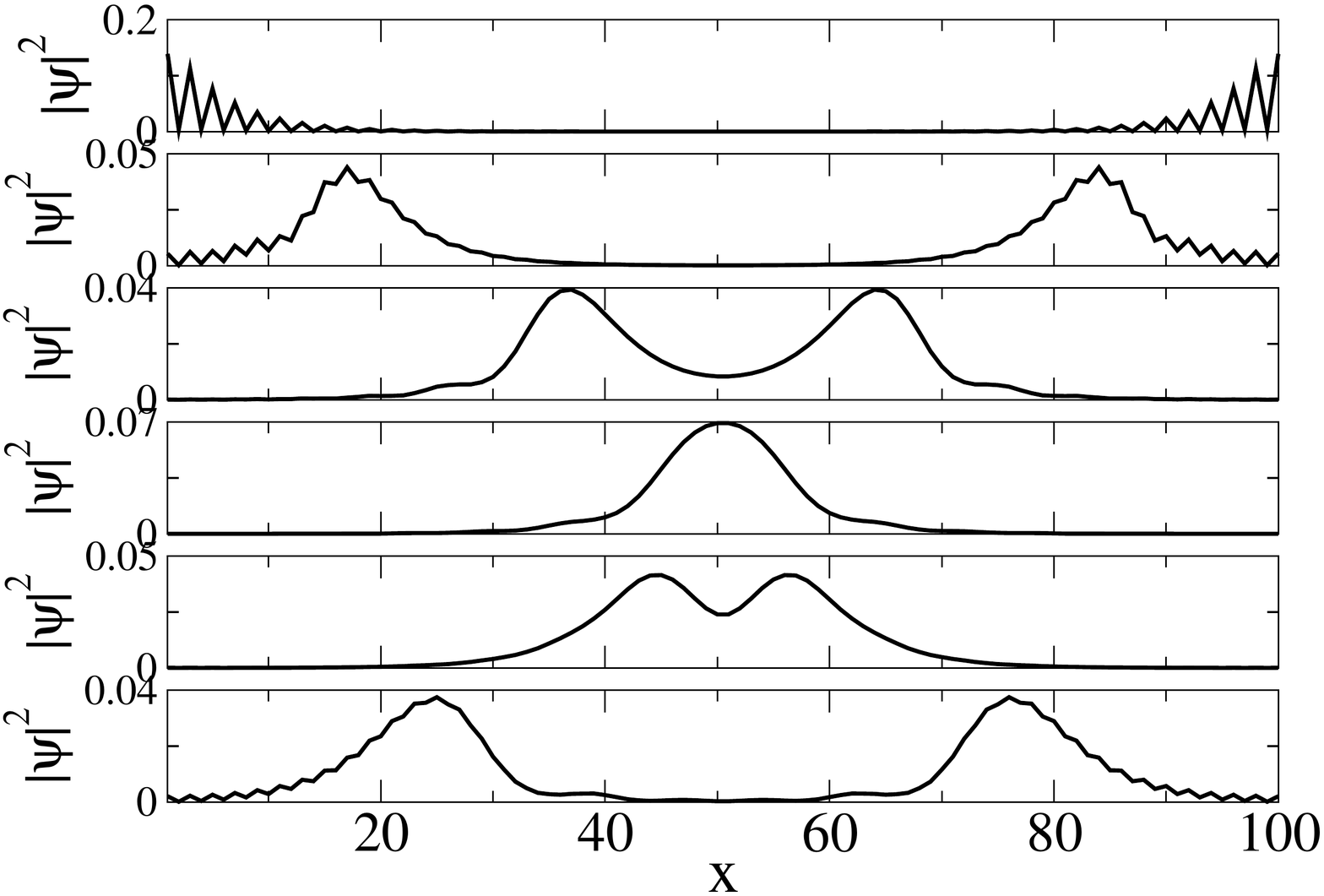}
\caption{\label{fig9}
Solitonic-like vs. constructive interference behavior in the Kitaev model. 
Top panel the initial state is far from the critical point (CP)
and in the lower panel one is close to the
CP.
}
\end{figure}

The origin of the period doubling is understood in the following way.
In Fig. \ref{fig9} the time evolution of the Majorana state is
shown for a critical quench from the region of oscillations, close to the critical point, and
a quench from a region where the period has doubled. In the first
case the wave functions at each edge are separated in two energy modes
while for the second they are mixed. This is due to the long range
correlations close to the critical point that effectively decrease the
system size and lead to the coupling of the two edge modes. In the first
case the time evolved states from each edge cross each other in a solitonic
like behavior while in the second case there is a constructive interference
when the peaks of the evolved state meet at the center of the wire.
Consistently with the results for the overlaps, in this regime the energy spectrum
between states with high weight halves, and the period doubles.

\begin{figure}
\includegraphics[width=0.85\columnwidth]{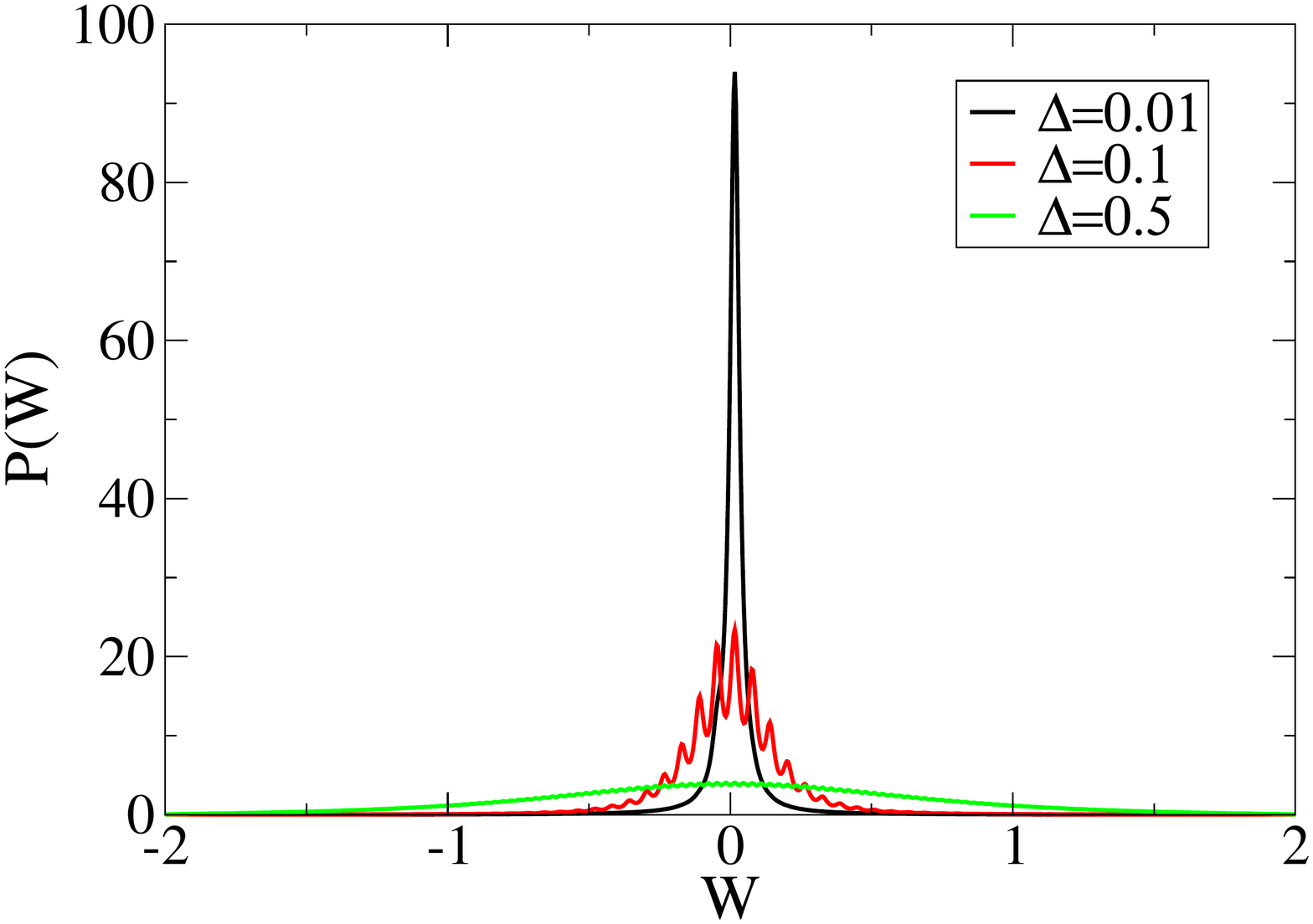}
\includegraphics[width=0.85\columnwidth]{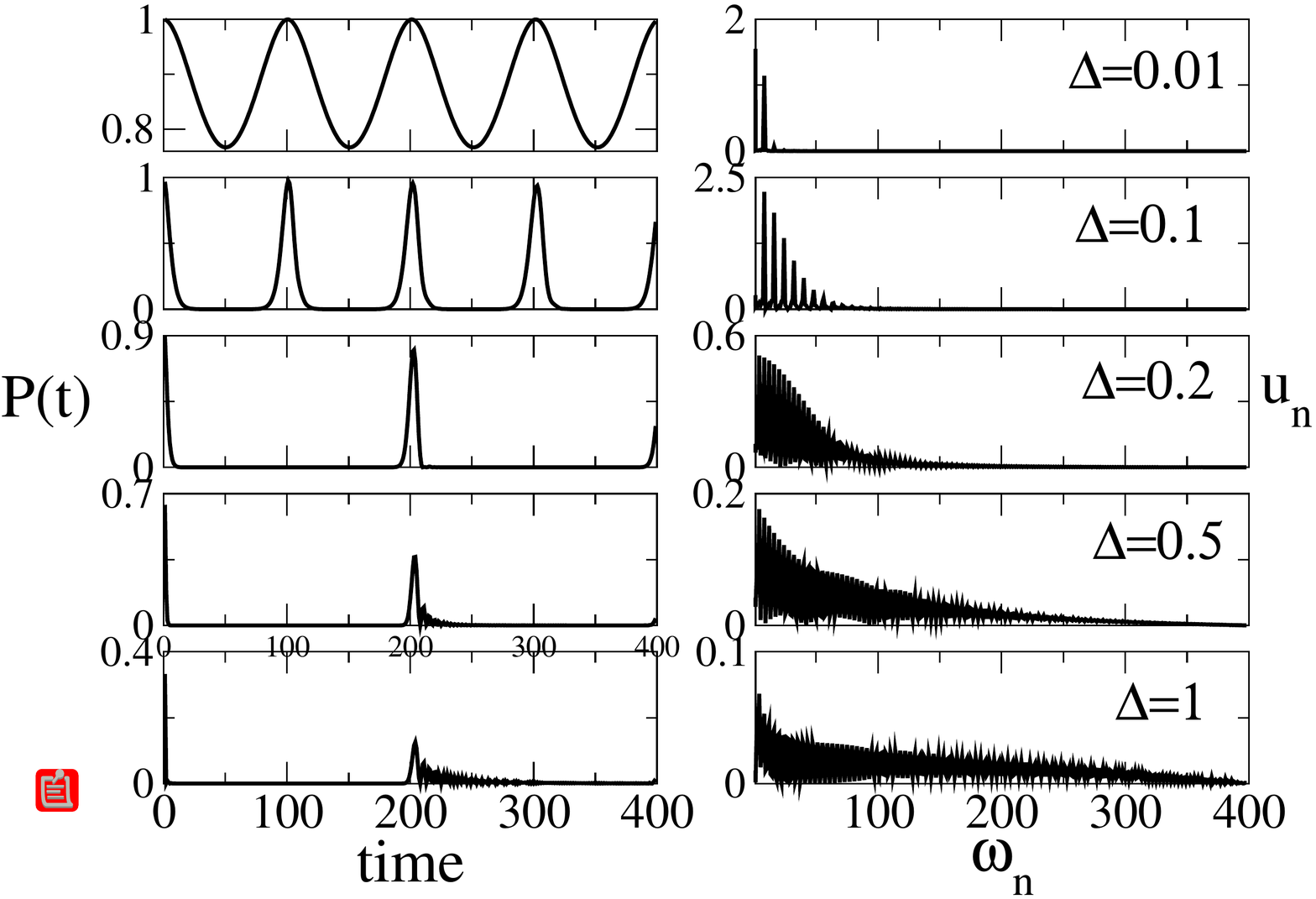}
\caption{\label{fig10}
(Color online) 
Top panel: Work distribution for the Kitaev model.
Lower panel: Fourier analysis for the Kitaev model. 
}
\end{figure}

The distribution of the overlaps may be parametrized by the
work distribution given by \cite{tasaki}
\be
P(W) = \sum_m P_m \delta (W-W_m)
\ee
where
\bea
P_m &=& |\langle \psi_m(\xi_1)|\psi_{m=0}(\xi_0) \rangle |^2 \nonumber \\
W_m &=& E_m(\xi_1)-E_{m=0}(\xi_0)
\eea
In Fig. \ref{fig10}a we show the work distribution associated with critical
quenches for $\mu=0$ to the critical point $\mu=0,\Delta=0$ starting 
from different initial points. 
Close to the critical point the
distribution is quite sharp but it becomes very broad as $\Delta$ increases.
The $\delta$-function peaks have been broadened for clarity.
A similar conclusion is obtained performing a Fourier
analysis of the time evolution of the survival probability. This is shown in the lower panel.
While for small initial values of $\Delta$ the distribution is quite narrow
around low frequencies,
it changes significanly as $\Delta$ grows, becoming quite extended.
In the Fourier decomposition the amplitudes, $u_n$, are for the frequencies with
values $\omega_n=\pi (n-1)/N_t$, where $N_t$ is the number of time points considered.

\begin{figure}
\includegraphics[width=0.75\columnwidth]{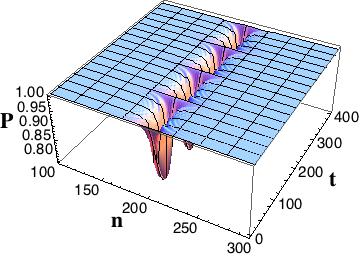}
\includegraphics[width=0.75\columnwidth]{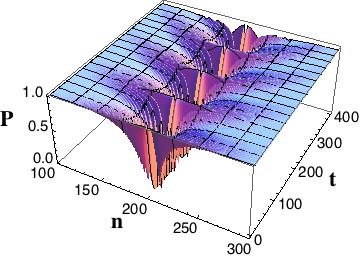}
\includegraphics[width=0.65\columnwidth]{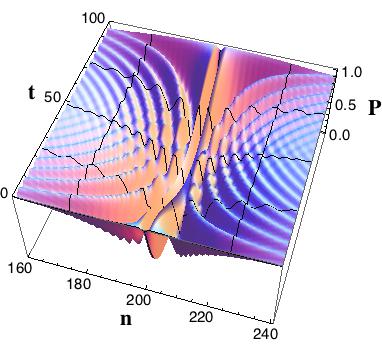}
\caption{\label{fig11}
(Color online) 
Survival probability of different initial single-particle
states, labeled by $n$ where $n=200,201$ are the Majorana zero energy modes. 
In the top panel the case of initial $\Delta=0.01$ is shown and in the middle panel
the case of $\Delta=0.1$ is shown. In the lower panel a low energy zoom is shown
of the case with $\Delta=0.1$.
}
\end{figure}

It is also interesting to study the survival probability of excited states,
that in this problem are
extended states throughout the chain. This is shown in
Fig. \ref{fig11}. Here we show the survival probability of different initial states, including
the Majorana states, for two cases of $\Delta=0.01,0.1$. 
Close to the critical point the survival probability of most states is close to $1$ except
near the low energy modes.  Further away from the critical point the deviation of the
survival probability from unity is larger due to larger orthogonality between the eigenstates
of the origin and final Hamiltonians. In the third panel the low energy region is enhanced
showing the complex behavior as a function of time and eigenstate, $n$.

\subsection{Generation of Majorana states}

\begin{figure}[h]
\includegraphics[width=0.75\columnwidth]{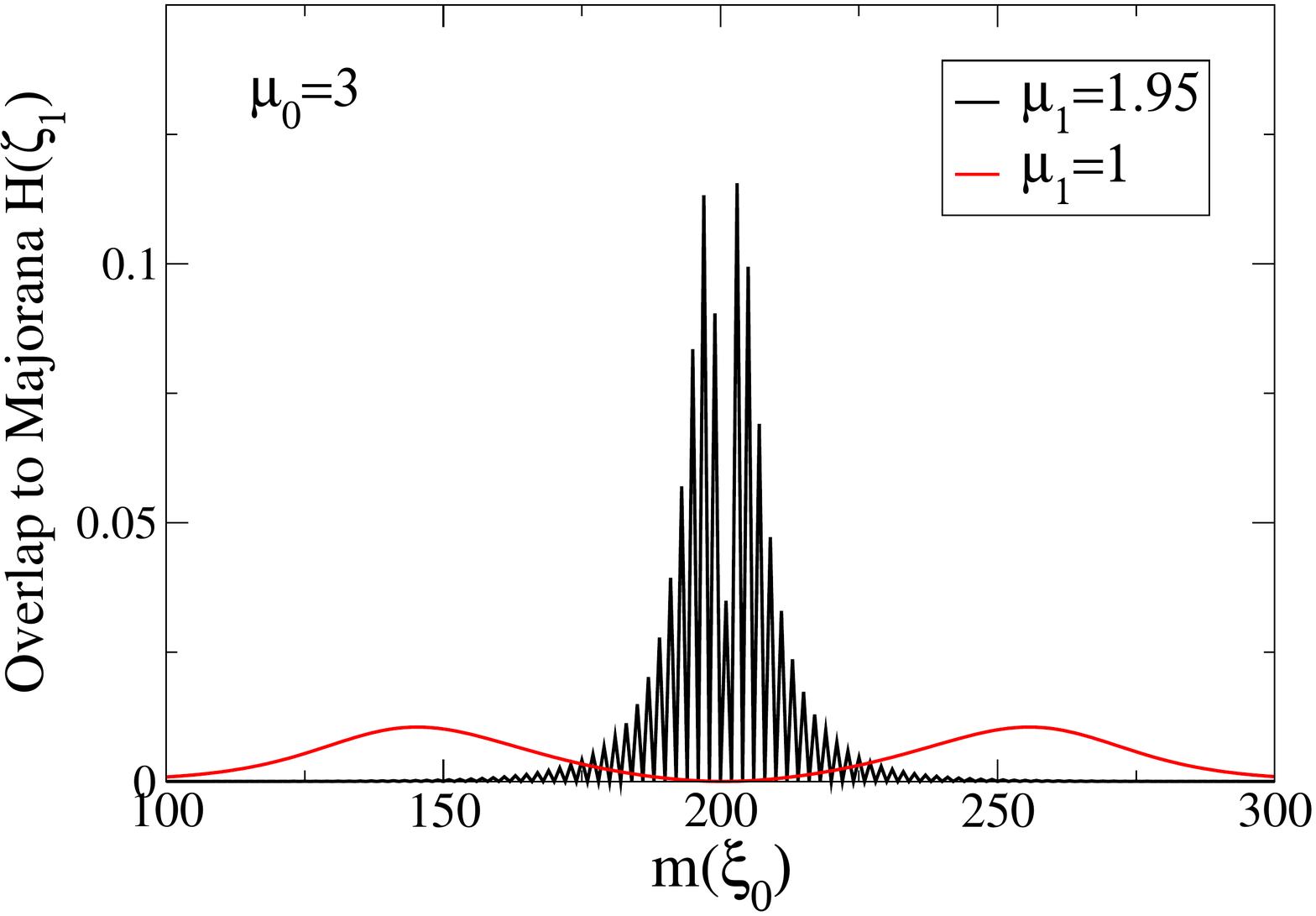}
\includegraphics[width=0.75\columnwidth]{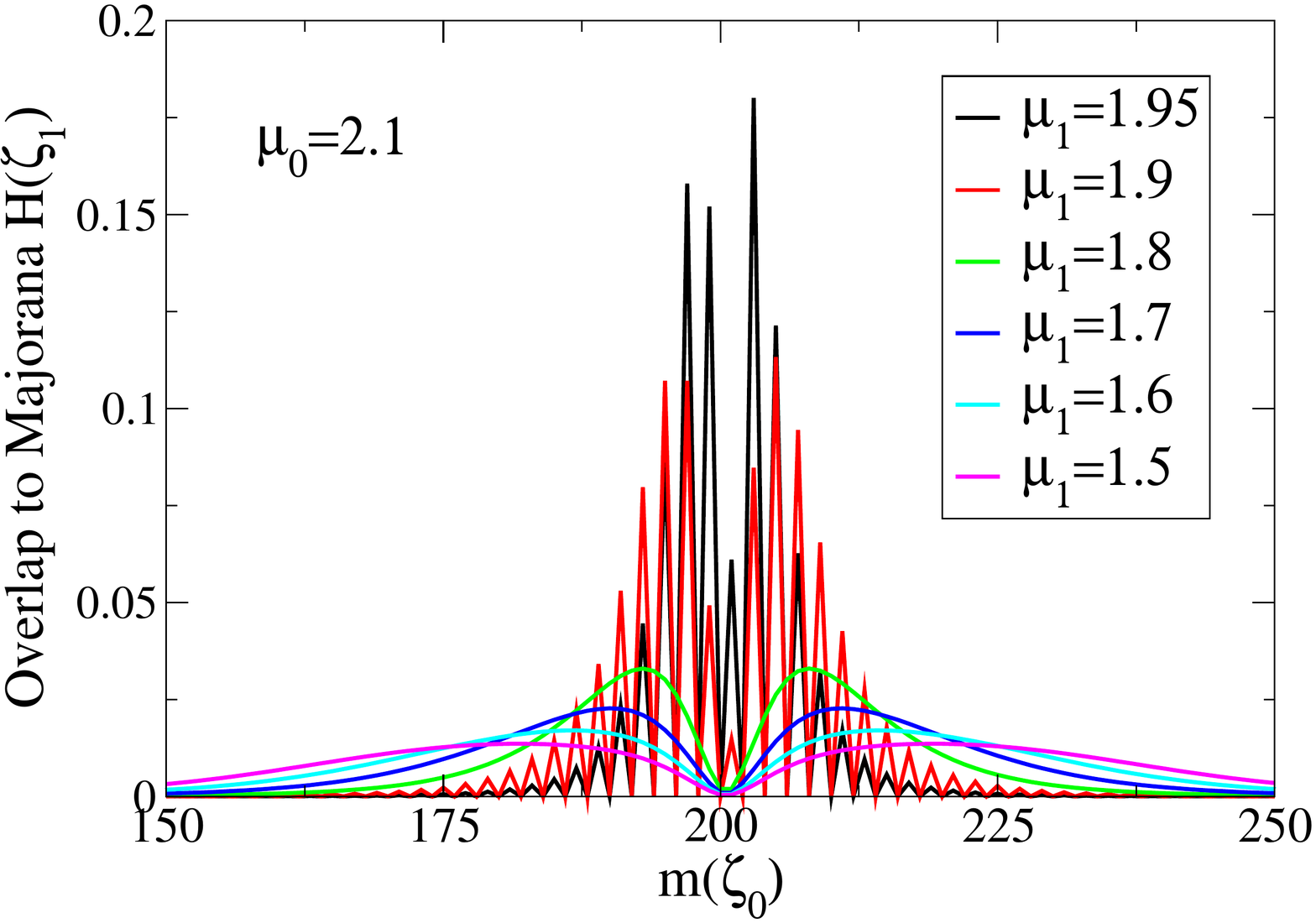}
\caption{\label{fig12}
(Color online) 
Overlap of states in trivial phase to Majorana final state in the topological phase far and
close to the critical line, respectively. 
}
\end{figure}

\begin{figure}
\includegraphics[width=0.75\columnwidth]{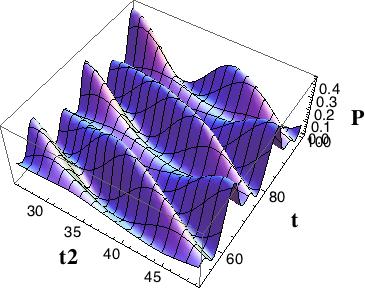}
\caption{\label{fig13}
(Color online) 
Probability to find the original Majorana state due to quench $\xi_2 \rightarrow \xi_3$
as a function of $t_2$ and $t$. The sequence of quenches is such that at $t_0=0$ there
is a quench $\xi_0 \rightarrow \xi_1$. At time $t_1$ there is a quench to $\xi_2=\xi_0$.
The time $t_1$ is chosen such that the survival probability of the original Majorana
vanishes. At time $t_2$ there is a new quench to $\xi_3$ in the topological region,
close to but different from $\xi_0$. The survival probability is finite and similar
results are obtained for other levels.  
}
\end{figure}

While quenches, either abrupt or slow, in general destabilize the edge states, topological phases 
can be induced by periodically driving the Hamiltonian
of a non-topological system.
The periodic driving leads to new topological states \cite{lindner}, and to a
generalization of the bulk-edge correspondence, that reveals a 
richer structure \cite{prx,balseiro} as
compared with the equilibrium situation \cite{delplace}, 
such as shown before in topological insulators \cite{oka,floquet0,lindner} 
and in topological superconductors, with the appearance of 
Majorana fermions \cite{jiang,luo,xiaosen,viola}.
Their appearance in a one-dimensional p-wave superconductor was studied in Ref. \cite{liu2}
and in Ref. \cite{manisha}; the case of
intrinsic periodic modulation was also considered \cite{foster}
and new phases may be induced and manipulated due to the presence of the 
periodic driving \cite{platero,hexagonal,liu2} or inducing spontaneous currents \cite{prb}.

Due to the finiteness of a system we may generate Majorana states through a sudden quench
starting from a
trivial phase. Even though, as stated above, in the thermodynamic limit the topological
properties can not be changed by a unitary transformation, as shown in
Fig. \ref{fig12} the probability that a given initial state in a trivial phase $III$
may collapse to a Majorana of the final state Hamiltonian in phase $I$ is finite. 
We have a finite although small overlap to the Majorana state of the topological phase.
This probability is given by
\be
|\langle \psi_{m_1=0}(\xi_1)|\psi_{m_0}^I(t) \rangle |^2 =
|\langle \psi_{m_1=0}(\xi_1)|\psi_{m_0}(\xi_0) \rangle |^2 
\ee
and is independent of time, as expected.
Quenching to a state close to the transition line, the overlaps of several (extended)
states are considerable due to the spatial extent of the Majorana states. If the quench is
deeper into the topological phase these become more localized and the overlap decreases.
Interestingly the larger overlap is found for higher energy, extended states.

A sequence of quenches allows for the manipulation of the states, as shown in eq. (9).
A possibility to turn off and on Majoranas can be trivialy seen in the following way.
Consider starting from a state inside region $I$. Perform a critical quench to the
line $\Delta=0$ and then a quench back to the original state. Choosing appropriately
$t_1$ we may get a state with no overlap with the initial Majoranas, as 
illustrated in Fig. \ref{fig6}.
So we are back to a topological phase but with no edge states. But Majoranas may
be switched back on if at a time $t_2>t_1$ we perform another quench to a state in
region $I$. This is illustrated in
Fig. \ref{fig13} where the survival probability is shown as a function of time and
intermediate time $t_2$ for a given time $t_1$ of the first quench. Due to the
quench to $\xi_3$ a finite probability to find the Majorana state is found even
though if no quench from $\xi_2=\xi_0 \rightarrow \xi_3$ was performed, and having
chosen appropriately $t_1$, the survival probability of the Majorana states was
tuned to vanish.
Note that the overlap of Majorana state of $H(\xi_3)$ with a Majorana state of
$H(\xi_0)$ is finite, since the states are chosen to be close by.

\section{Dynamics of multiband systems}

While in the previous sections Majorana edge states were considered, edge states
in other systems, including topological insulators, have also been considered and
show similar properties. 
In this section we consider two topological systems, the Schockley model \cite{yakovenko}
which has fermionic edge states and no Majoranas, and the SSH-Kitaev model \cite{tanaka}
which displays both types of edge states in different parts of the phase diagram,
allowing a comparison of different edge states.
Similarities and differences will be addressed and oscillations are also seen due to
off-critical quenches (accross a quantum phase transition) similarly to those
seen in the Kondo model \cite{henrik}, and due to edge states associated with high-energy gaps
in the spectrum, due to the multiband structure of these models.

\subsection{Schockley model}

\begin{figure}
\includegraphics[width=0.85\columnwidth]{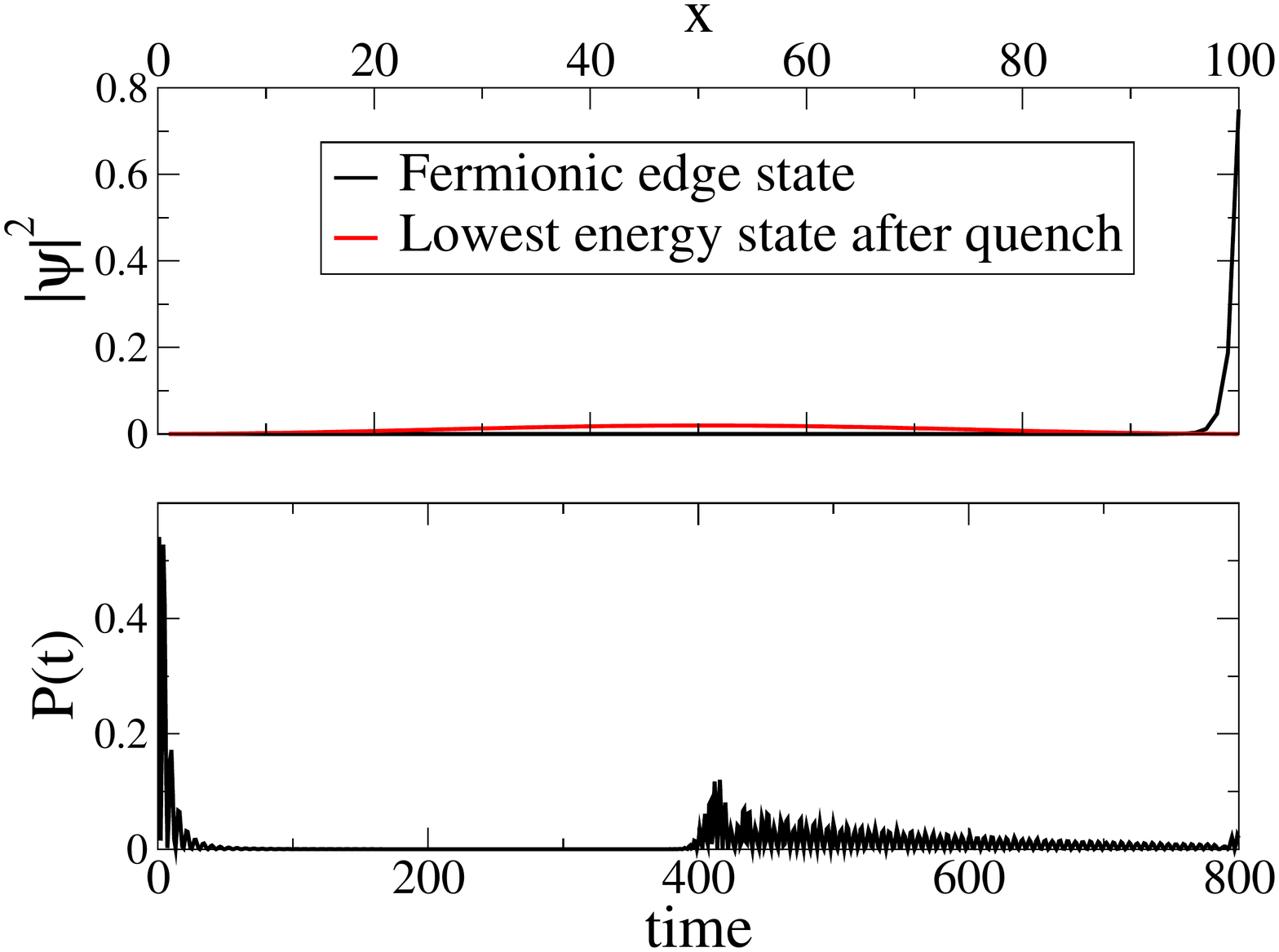}
\includegraphics[width=0.85\columnwidth]{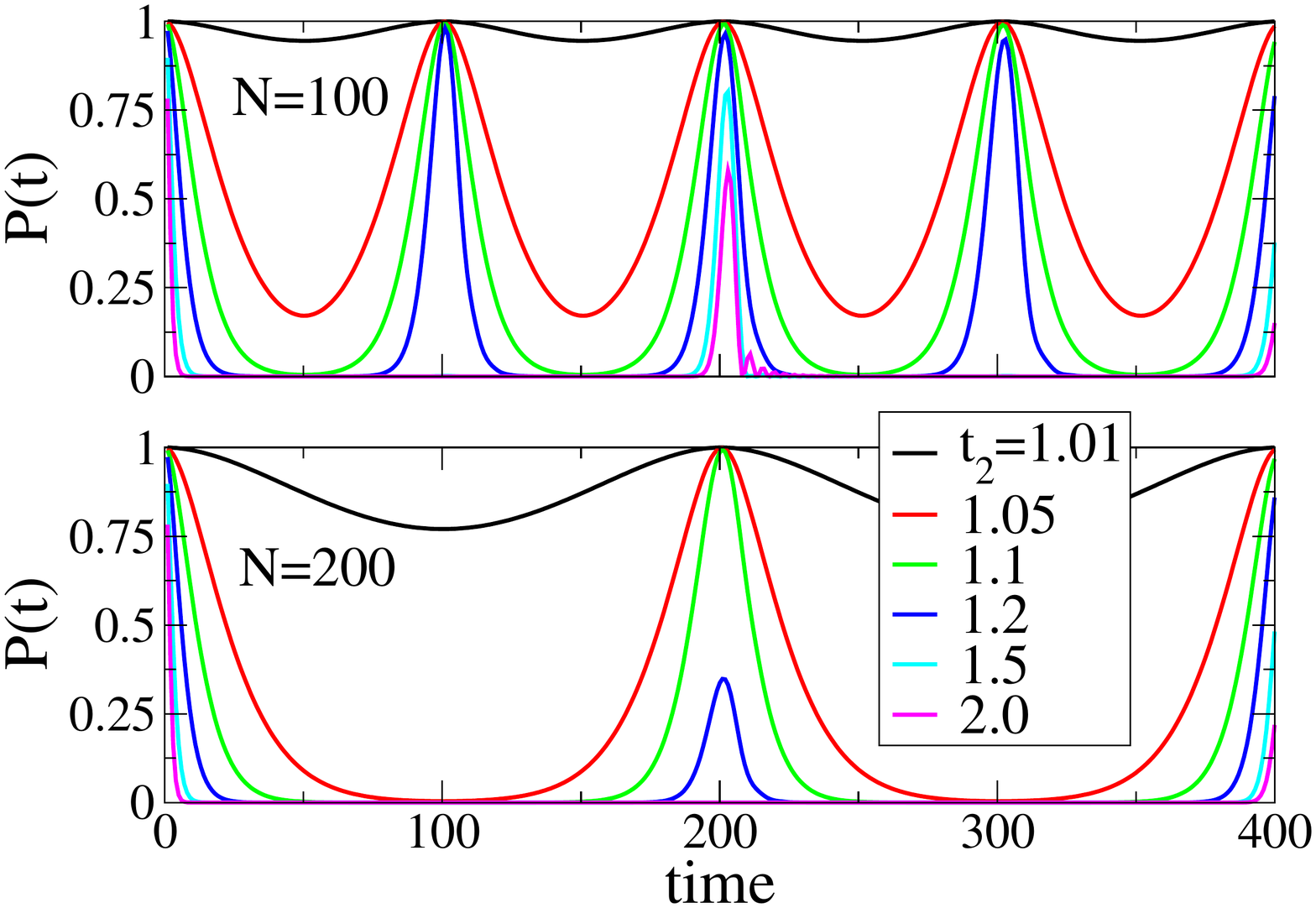}
\caption{\label{fig14}
(Color online) 
Survival probability of edge modes of Schockley model.
Top panel: off-critical quench from the topological region 
($t_1=1,t_2=2$) to the trivial region ($t_1=1,t_2=0.5$). In
the lower panel critical quenches are considered from the
topological region to the transition point ($t_1=t_2$).
}
\end{figure}

In Fig. \ref{fig14}a is considered a off-critical quench from the topological phase
with $t_1=1, t_2=2$ to the trivial phase $t_1=1,t_2=0.5$. In the first
panel the {\it fermionic} (not Majorana) 
very localized state of the initial phase and the lowest energy state
of the final state wave functions are shown. In the lower panel
the survival probability is shown. It is very similar to the case of
a non-critical quench in the Kitaev model with a rapid decrease of the
survival probability, and a revival (with small amplitude) due to the finiteness of the system,
with many frequenciey modes contributing to the dynamics, as characteristic of a quench far
from the critical point ($t_1=t_2$).

In Fig. \ref{fig14}b we show critical quenches to a final state with
$t_2=t_1=1$ starting from different initial points in the topological
region ($t_2>t_1$). The cases of $N=100$ and $N=200$ are shown. As
in the Kitaev model the period scales with the system size. The behavior
is very similar to the Kitaev model. We see the period doubling for
both cases for $t_2=1.5$.
For $t_2=2.0$ the smoothness of the oscillations is replaced by a superposition
of many frequencies.
From the point of view of edge state dynamics the behavior of Majoranas and fermionic
edge states are similar.

\subsection{SSH-Kitaev model}

\begin{figure}
\includegraphics[width=0.75\columnwidth]{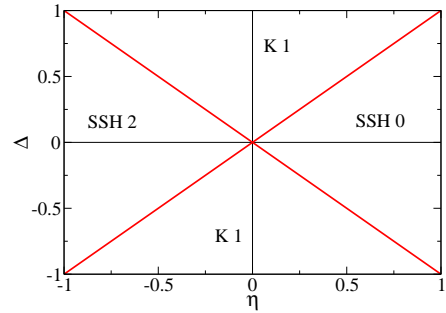}
\caption{\label{fig15}
(Color online) 
Phase diagram of SSH-Kitaev model for $\mu=0$. 
}
\end{figure}

The similarities are further shown considering the SSH-Kitaev model.
In Fig. \ref{fig15} we show the phase diagram of the SSH-Kitaev model \cite{tanaka}
in the case of $\mu=0$. 
In phase K1 we are in the Kitaev regime with one zero energy edge mode at each edge
(Majoranas). In the SSH regimes we are closer to the behavior of the SSH model with
fermionic modes. In SSH 0 there are no edge modes. In SSH 2 there are two zero energy
fermionic modes.

\begin{figure}
\includegraphics[width=0.75\columnwidth]{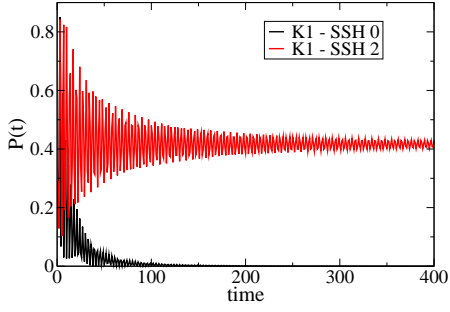}
\caption{\label{fig16}
(Color online) 
Off-critical quenches for the SSH-Kitaev model accross a transition line.
The parameters are $\mu=0, \eta=0, \Delta=0.2$ to $\eta=0.7$ and $\eta=-0.7$, respectively. 
}
\end{figure}

In Fig. \ref{fig16} we consider off-critical quenches in the
SSH model. 
In this figure we compare two quenches one from K1 to SSH 0 and the other to
SSH 2.
There are no oscillations. In the first case there are no matching states in the
final state and in the second case there is a finite overlap even though the initial
state is a Majorana and the final state has fermionic modes. Note that in the Kitaev (K1)
regime there is one Majorana at each edge and in SSH 2 there are two fermionic modes
(four Majoranas coupled two by two to form fermionic modes) at each edge.
This is similar to the Kitaev model in the sense that quenches to the trivial
region lead to a vanishing survival probability after a short time, and a transition
from a topological phase to another point in the topological phase leads to a
finite probability. Note that in the Kitaev model a transition between the two
topological regions $I$ and $II$ leads to a vanishing survival probability due
to the orthogonality of the edge states \cite{rajak}.

\begin{figure}
\includegraphics[width=0.75\columnwidth]{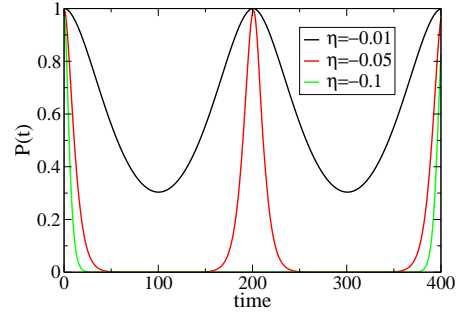}
\includegraphics[width=0.75\columnwidth]{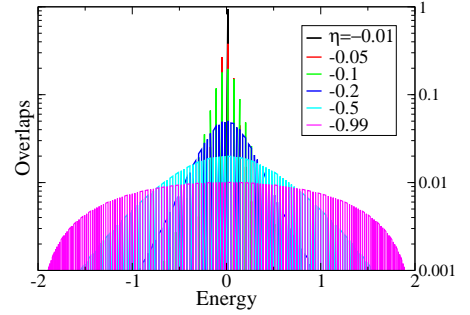}
\includegraphics[width=0.75\columnwidth]{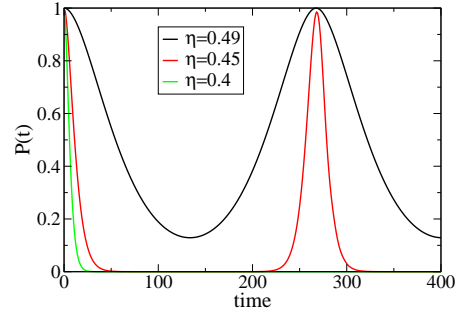}
\includegraphics[width=0.75\columnwidth]{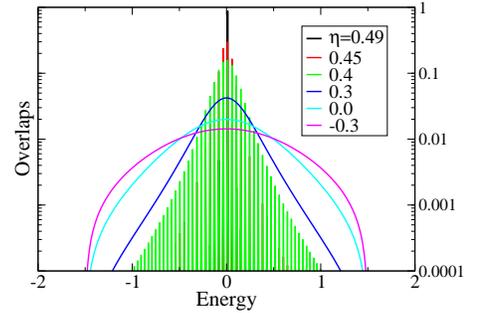}
\caption{\label{fig17}
(Color online) 
Critical quenches in the SSH-Kitaev model: survival probability and overlaps.
The CP on the first two panels is $\eta=0,\Delta=0$ (SSH2) and on the last two panels the CP is
$\eta=0.5, \Delta=0.5$ (K1).  
}
\end{figure}

In Fig. \ref{fig17} 
we consider critical quenches to points in the transition between different
topological regions. In the top panels we consider $P(t)$ and the overlaps of a transition
at $\mu=0$ from the SSH 2 regime to the critical point $\eta=0, \Delta=0$ by considering
different initial values of $\eta=-0.01,-0.05,-0.1,-0.2,-0.5,-0.99$. In the lower
panels we consider critical quenches to the critical point $\eta=0.5, \Delta=0.5$ changing
the initial value of $\eta$. In both cases note that there is again a change of the distribution
of the overlaps from sharp peaks, at small deviations from the critical point,
to a broad distribution of the overlaps as one moves sufficiently away from the
critical point; again there is a crossover between the two regimes (not shown), 
as for the Kitaev model. However, the overlaps are not smooth as a function of energy.
Note that in the first case $\Delta=0$ which means this occurs in the context of the
SSH model with no superconductivity. In the second case we have a mixture of SSH and
Kitaev model but the behavior is qualitatively similar in the crossover region. Beyond
it we find again the very smooth distributions of the overlaps as in the Kitaev model.

\begin{figure}
\includegraphics[width=0.75\columnwidth]{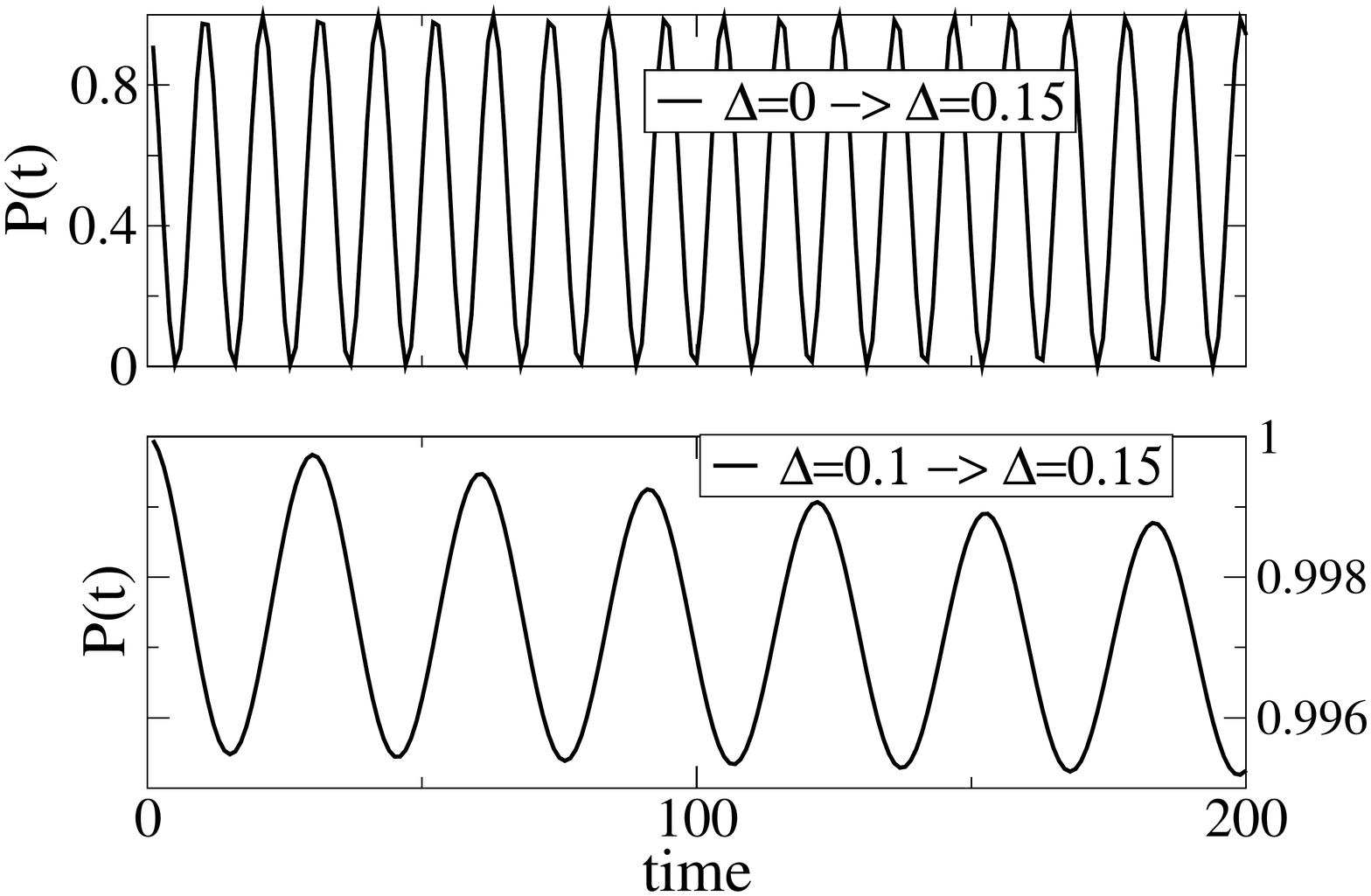}
\includegraphics[width=0.75\columnwidth]{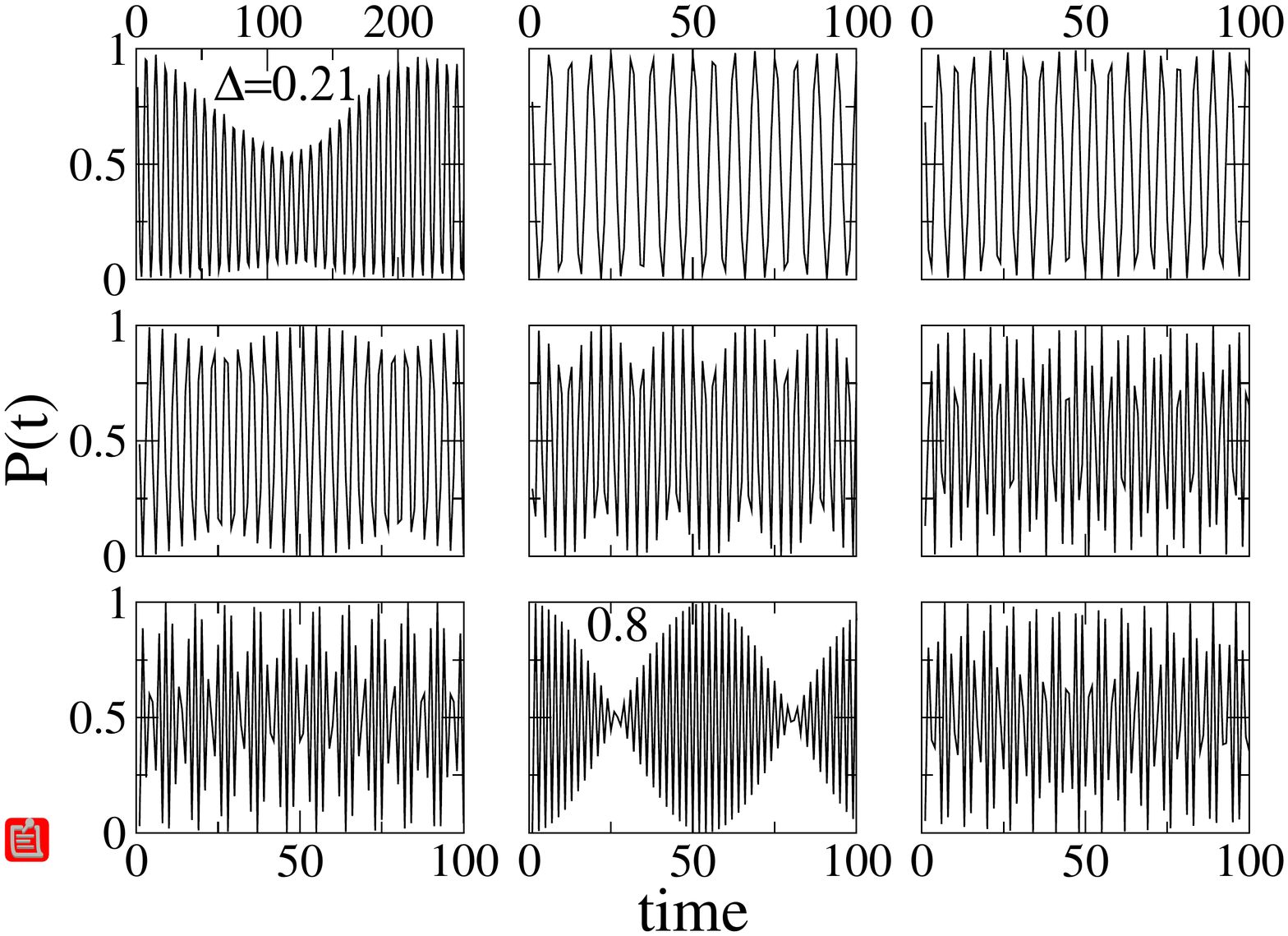}
\caption{\label{fig18}
SSH0-K1 transition in the SSH-Kitaev model (trivial to topological). Originally we have
an extended state and in the final we have an edge state but not very localized.
In the top panel we have $\mu=0$ and $\eta=0.7, \Delta=0$ to $\eta=0.1, \Delta=0.15$
or the initial value of $\Delta=0.1$.
In the lower panel we have also SSH0 to K1 and $\mu=0$ but we start from
$\eta=0.2, \Delta=0$ and change to $\Delta=0.21,0.25,0.3,0.4,0.5,0.6,0.7,0.8,0.9$.
}
\end{figure}

In Fig. \ref{fig18} 
in the top panel the
parameters are  $\mu=0, \eta=0.7, \Delta=0$ or $\Delta=0.1$ that get
changed to $\mu=0, \eta=0.1, \Delta=0.15$. 
There are oscillations but if 
$\Delta_0=0.1$ oscillations are also observed but there is
also a noticeable decay of the amplitude of the survival probability.
In the lower panel is considered an initial state with $\mu=0, \eta=0.2, \Delta=0$
and various final states with different values of $\Delta$. 
Again there are
oscillations in some regimes with an admixture of other frequencies as $\Delta$
changes. 
Note that we are considering the survival probability of a trivial extended state, since we
start from the phase SSH0. The quench takes place to a topological phase with Majorama edge
states but the overlaps are summed over all eigenstates, and so one expects a finite $P(t)$.
However, the presence of oscillations is still significative, although several frequencies
contribute, as seen by the modulation of the oscillations.

\begin{figure}
\includegraphics[width=0.75\columnwidth]{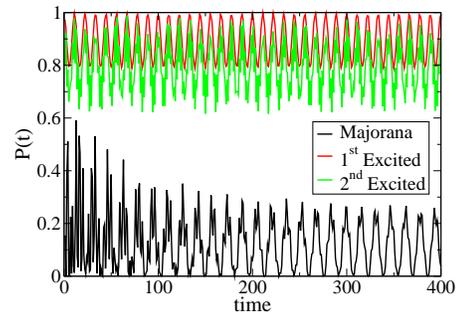}
\caption{\label{fig19}
(Color online) 
Transition K1-SSH0 in the SSH-Kitaev model. 
}
\end{figure}

In Fig. \ref{fig19}
we consider a transition from K1 to SSH0 but with $\mu=0.2$ (see Ref. \cite{tanaka} for
phase diagram).
The initial state has one edge Majorana but the final state, even though
$\eta$ is negative $\eta=-0.7$, there are no edge zero energy fermionic modes.
If $\mu=0$ there are modes with zero energy but if the chemical potential is finite
these modes have finite energy. But they are localized at the edge of the chain.
This is like in ref. \cite{rio}. 
In the figure is compared the survival probability of the lowest energy state (Majorana
state of the K1 phase; note that in the Kitaev regime even though the chemical potential
does not vanish the edge state has zero energy and therefore is a Majorana). If $\Delta=0$
and $\mu\neq 0$ then there is a possibility of edge states with localized wave functions
but finite energy. If $\mu=0$ then these fermionic modes have zero energy, as discussed
above. The behavior is a bit complex.

\begin{figure}
\includegraphics[width=0.75\columnwidth]{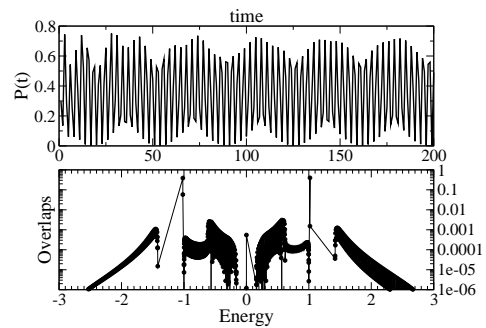}
\caption{\label{fig20}
Modulated frequencies and finite energy peaks in the SSH-Kitaev model. 
The transition is of the type SSH2-K1.
}
\end{figure}

In Fig. \ref{fig20} 
we show another example of a modulated frequency due to mixture of finite
overlaps as shown in the bottom panel. The parameters here are
$\mu=0, \eta=-0.2, \Delta=0.1$ to $\mu=1, \eta=-0.2, \Delta=0.1$.
In this case there is a transition from a state with two zero energy edge states
to a state with one zero energy edge state. Note that there are large
overlaps to states that are at the edge of the high energy gap.
These states are high-energy localized states and therefore with a somewhat spatial distribution
as the Majorana or fermionic edge states at low energies.

\begin{figure}
\includegraphics[width=0.9\columnwidth]{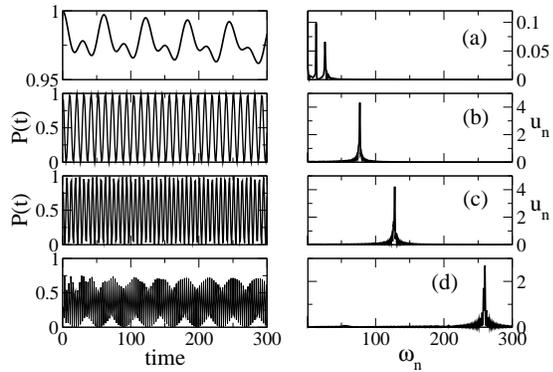}
\caption{\label{fig21}
Fourier decomposition of several quenches in the SSH-Kitaev model. 
The parameters are: a) $\mu=0,\eta=0.7,\Delta=0.15 \rightarrow \mu=0,\eta=0.1,\Delta=0.15$ (SSH0 $\rightarrow $ K1), 
b) $\mu=0,\eta=0.7,\Delta=0 \rightarrow \mu=0,\eta=0.1,\Delta=0.15$ (SSH0 $\rightarrow $ K1),
c) $\mu=0,\eta=0.2,\Delta=0 \rightarrow \mu=0,\eta=0.2,\Delta=0.25$ (SSH0 $\rightarrow $ K1),
d) $\mu=0,\eta=-0.2,\Delta=0.1 \rightarrow \mu=1,\eta=-0.2,\Delta=0.1$ (SSH2 $\rightarrow $ K1).
}
\end{figure}

As shown in Fig. \ref{fig21} by the Fourier decomposition of the time dependence
of $P(t)$ for various quenches, important contributions may be due to finite energy states.

\section{Conclusions}

Topological systems are robust to unitary transformations in the thermodynamic limit.
However, finite systems and their associated edge states are in general not robust.
In this work the dynamics of these edge modes is analysed with particular emphasis
on the oscillations of the survival probability of a single-particle state after a
sudden quench of the Hamiltonian parameters.

Majorana and fermionic zero-energy modes were compared and their general behaviors
are similar. Differences occur mainly due to the specifics of each transition between
different phases of the various topological systems.
The survival probability is controlled by the overlaps between the eigenstates of the
Hamiltonians prior and after the quantum quench, as well as by the excitation spectrum
of the final state Hamiltonian.
While transitions between points in parameter space in the same topological phase or
between points in different phases (off-critical quenches) have been studied before,
here we have focused on critical quenches, where often oscillations in the 
survival probability, $P(t)$, appear. It turns out that oscilations 
also occur in some off-critical quenches.

The regime of oscillations, or more loosely periodicity of $P(t)$, is changed as the initial
state approaches the critical region. Specifically, even-odd effects or their absence lead to
a period doubling, whose crossover depends on the system size (in a way similar to 
the revival time
scaling previously considered). The critical fluctuations near a critical point (or line
of points) effectively decrease the system size, leading to a more pronounced coupling of the
states at the two edges of the system.
These considerations hold both for the Majorana edge states, found in topological superconductors
(exemplified here by the $1d$ Kitaev model), and for the fermionic zero-energy states of
a topological insulator (exemplified here by Schockley model). An interesting model
that provides both Majorana and fermionic edge states is the SSH-Kitaev model considered here.
Its multiband structure also reveals interesting oscillation effects due to the presence
of large overlaps to finite energy states (appearing in high-energy gaps of the spectrum)
since these states are also localized. There is no clear-cut distinction between the
various states localized at the edges of the system, from the point of view of their
contribution to the dynamics of the survival probability.

Pushing further the consequences of the finiteness of the system it is trivial to find cases
where Majoranas can be generated by the dynamical process, in the sense that the overlap
between single-particle states of the trivial phase of Kitaev model and a Majorana state
of the topological regime is in some cases finite, and moreover time independent.
Also, one may switch off and back on Majorana states if sequences of quenches are chosen
appropriately, as exemplified in the text.

\vspace{\baselineskip}

{\bf Acknowledgements}

The author acknowledges several discussions with Henrik Johannesson and partial
support and hospitality
by the Department of Physics of Gothenburg University, where most of this work
was carried out. A question by Rina Takashima at the Department of Physics of Kyoto University
lead to part of this work. Partial support from FCT through grant UID/CTM/04540/2013 
is also acknowledged.

\end{document}